\documentclass[12pt]{article}
\usepackage{amsfonts}
\usepackage{latexsym}
\usepackage{amsmath,amssymb}
\usepackage{verbatim}
\usepackage{setspace}
\usepackage{color}
\usepackage{physics}
\usepackage{tikz}
\usepackage{tikz-cd}
\usepackage{mathdots}
\usepackage{cancel}
\usepackage{color}
\usepackage{siunitx}
\usepackage{array}
\usepackage{multirow}
\usepackage{amssymb}
\usepackage{gensymb}
\usepackage{tabularx}
\usepackage[normalem]{ulem}
\usepackage{booktabs}
\usetikzlibrary{fadings}
\usetikzlibrary{patterns}
\usetikzlibrary{shadows.blur}
\usetikzlibrary{shapes}
\usepackage{cite}
\usepackage{hyperref}
\usepackage[mathscr]{euscript}
\numberwithin{equation}{section}

\newcommand{\p}{\partial}

\newcommand{\bit}{\begin{itemize}}
\newcommand{\eit}{\end{itemize}}
\newcommand{\bd}{\begin{description}}
\newcommand{\ed}{\end{description}}

\newcommand{\bc}{\begin{center}}
\newcommand{\ec}{\end{center}}

\newcommand{\be}{\begin{equation}}
\newcommand{\ee}{\end{equation}}
\newcommand{\bea}{\begin{eqnarray}}
\newcommand{\eea}{\end{eqnarray}}
\newcommand{\bs}{\begin{subequations}}
\newcommand{\es}{\end{subequations}}

\def\p{\partial}

\def\bz{\bar{z} }

\usepackage{cite}
\usepackage{hyperref}

\begin{document}

\begin{titlepage}

\unitlength = 1mm
\ \\
\vskip 2cm
\begin{center}
{\LARGE{\textsc{Celestial Sector in CFT: \vspace{12pt}\\
Conformally Soft Symmetries}}}

\vspace{0.8cm}
Leonardo Pipolo de Gioia$^1$ and Ana-Maria Raclariu$^2$
\vspace{0.7cm}\\
\small{${}^{1}$\textit{Institute of Physics ``Gleb-Wataghin'', University of Campinas - UNICAMP}, \\
\textit{13083-859, Campinas - SP, Brazil}
\vspace{5pt}\\
${}^2$\textit{Institute of Physics, University of Amsterdam, Science Park 904,\\ 1198 XH, Amsterdam, the Netherlands}} 
\vspace{20pt}

\begin{abstract}

We show that time intervals of width $\Delta \tau$ in 3-dimensional conformal field theories (CFT$_3$) on the Lorentzian cylinder admit an infinite dimensional symmetry enhancement in the limit $\Delta \tau \rightarrow 0$. The associated vector fields are approximate solutions to the conformal Killing equations in the strip labelled by a function and a conformal Killing vector on the sphere. An Inonu-Wigner contraction yields a set of symmetry generators obeying the extended BMS$_4$ algebra. We analyze the shadow stress tensor Ward identities in CFT$_d$ on the Lorentzian cylinder with all operator insertions in infinitesimal time intervals separated by $\pi$. We demonstrate that both the leading and subleading conformally soft graviton theorems in $(d-1)$-dimensional celestial CFT (CCFT$_{d-1}$) can be recovered from the transverse traceless components of these Ward identities in the limit $\Delta \tau \rightarrow 0$. A similar construction allows for the leading conformally soft gluon theorem in CCFT$_{d-1}$ to be recovered from shadow current Ward identities in CFT$_d$.

\end{abstract}
\vspace{0.5cm}
\end{center}
\vspace{0.8cm}

\end{titlepage}
\tableofcontents

\section{Introduction}

Celestial holography proposes a correspondence between theories of gravity in 4-dimensional (4D) asymptotically flat spacetimes and conformal field theories (CFT) living on the 2D celestial sphere at infinity \cite{Pasterski:2016qvg, Pasterski:2021raf}. In particular, scattering observables in the 4D theory are computed by correlation functions in the 2D theory, also known as celestial amplitudes,\footnote{Celestial amplitudes will be assumed to be defined in 2D whenever the dimension is not explicitly specified.} and are subject to a wide range of symmetries \cite{Stieberger:2018onx, Donnay:2018neh, Pate:2019mfs, Nandan:2019jas, Adamo:2019ipt, Puhm:2019zbl, Pate:2019lpp, Donnay:2020guq} (see also \cite{McLoughlin:2022ljp} for a recent review). This correspondence appears to be very different from other instances of holography. Most notably, it relates a bulk theory to a boundary theory in \textit{two} lower dimensions, while the bulk soft theorems imply the existence of towers of negative dimension operators in the celestial CFT \cite{Guevara:2019ypd}, naively rendering the boundary theory non-unitary.  

On the other hand, for massless\footnote{It has been long known that massive and in some cases massless momentum space scattering amplitudes can be extracted from correlation functions of unitary CFT$_d$ with holographic AdS$_{d+1}$ duals in various flat space limits \cite{Polchinski:1999ry, Giddings:1999jq, Penedones:2010ue, Paulos:2016fap, Paulos:2017fhb}. Interestingly, it was recently shown that such CFT$_d$ 4-point correlators exhibit conjectured properties of $(d+1)$-dimensional scattering amplitudes, including dispersion relations, unitarity and the Froissart bound in a flat-space limit \cite{vanRees:2022itk}.} scattering, a simple flat space limit of holographic CFT$_d$ correlators was found in \cite{PipolodeGioia:2022exe} to yield  $(d - 1)$-dimensional celestial amplitudes. This suggests that at least some of the unique features of celestial CFT should arise in a certain limit of conventional CFT in one higher dimension. The goal of this paper is to explain how leading and subleading conformally soft symmetries \cite{Pate:2019mfs,Nandan:2019jas, Adamo:2019ipt, Puhm:2019zbl} emerge precisely in this way. 

Motivated by the configuration of boundary operators for which CFT$_3$ correlators reduce to celestial amplitudes, we first study the symmetries of an interval on the Lorentzian cylinder of small width $\Delta \tau \propto R^{-1}$ in global time. We show that in the limit $R \rightarrow \infty$, the conformal isometries of this strip are enhanced to an infinite dimensional symmetry parameterized by a function and a local conformal Killing vector on a two-sphere. For finite large $R$ (corresponding to a strip of small, but finite width), the infinite dimensional symmetry is broken by $O(R^{-1})$ terms. We show explicitly via a procedure that mimics the Inonu-Wigner contraction \cite{Inonu} of the conformal algebra to Poincar\'e, that the enhanced conformal isometries of the intervals around $\tau = \pm \frac{\pi}{2}$ generate an extended BMS$_4$ algebra to leading order at large $R$. Moreover, under these symmetries, CFT$_3$ primary operators of dimension $\Delta$ at $\tau = \pm\frac{\pi}{2}+ \frac{u}{R}$ transform as 2D primary operators of effective dimension $\hat{\Delta} = \Delta + u\p_u$. $\hat{\Delta}$ can be diagonalized by an integral transform with respect to $u$ analogous to that relating Carrollian and celestial operators \cite{Donnay:2022aba, Donnay:2022wvx}. 

This analysis suggests that conformally soft symmetries in 2D CCFT are generated by certain modes of the 3D stress tensor in the strips. In the second part of the paper we show that the \textit{shadow} stress tensor Ward identities in CFT$_d$ allow one to extract both the leading and subleading conformally soft graviton operators in CCFT$_{d-1}$. We establish this by lifting the method used in \cite{Kapec:2017gsg} to derive stress tensor Ward identities from the subleading soft graviton theorem in arbitrary dimensions to the embedding space. This allows us to derive the shadow stress tensor Ward identities on the Lorentzian cylinder $\mathbb{R} \times S^{d-1}$ and study their restriction to an infinitesimal global time strip. Specifically, we find that 
\be 
\underset{u \rightarrow 0}{\lim} \p_u \widetilde{T}_{ab} ~~ {\rm and} ~~ \underset{u \rightarrow 0}{\lim} (1 - u \p_u)\widetilde{T}_{ab},
\ee
where $\widetilde{T}_{ab}$ is the shadow transform of the CFT$_d$ stress tensor and $a,b$ are indices on $S^{d-1}$ become respectively, upon subtracting the trace, the leading and subleading conformally soft gravitons in CCFT$_{d-1}$!

Our results are interesting for several reasons. Firstly, they demonstrate that celestial CFT may not be as exotic of a theory as anticipated. On the contrary, the leading and subleading conformally soft symmetries arise universally in a simple limit of any CFT$_3$, irrespective of whether or not it is holographic. In this sense, our approach is complementary to that in \cite{Hijano:2020szl, Banerjee:2021llh, Banerjee:2022oll} which relies on the existence of an AdS bulk dual. More generally, we find that any CFT$_{d}$ contains a $(d - 1)$-dimensional ``celestial'' sector characterized by an emergent BMS-like symmetry.\footnote{In $d > 3$ the vector fields are parameterized by a function on the sphere and a CKV on $S^{d-1}$, in particular there is no local enhancement of the latter like for $d = 3.$} Secondly, our results suggest that holographic CFT$_d$ correlators encode information about gravity in $(d+1)$-dimensional asymptotically flat spacetimes (AFS) that need not be lost in the flat space limit. It would be extremely interesting to understand the further implications, as well as the limitations of this approach.

This paper is organized as follows.  In section \ref{sec:preliminaries} we review the relation between AdS Witten diagrams and celestial amplitudes at large AdS radius. We show how each operator in an infinitesimal time interval around $\tau = \pm \frac{\pi}{2}$ in a CFT$_d$ on the Lorentzian cylinder maps to a continuum of operators in CCFT$_{d-1}$ via an integral transform over the interval. In section \ref{sec:spin} we generalize the relation between AdS Witten diagrams and celestial amplitudes to massless spinning external states. In particular, we demonstrate that, at large AdS radius, spinning bulk-to-boundary propagators in AdS$_{d+1}$ with fixed dimensions become massless spinning conformal primary wavefunctions in $\mathbb{R}^{1,d}$. In section \ref{sec:infinite-dimensional} we analyze the conformal Killing equations in a global time strip of the 3D Lorentzian cylinder of infinitesimal width $\Delta \tau \sim R^{-1}$. We find an emergent infinite dimensional symmetry in limit $R \rightarrow \infty$ labelled by a function and a vector field on the sphere. We show in section \ref{subsec:extendedBMS} that the associated vector fields reorganize into the generators of an extended BMS$_4$ algebra after a Inonu-Wigner-like contraction. In section \ref{sec:ckv-action} we show that CFT$_3$ operators in the strips around $\tau = \pm \frac{\pi}{2}$ transform like conformal primary operators in CCFT$_{2}$ under these symmetries.

In section \ref{sec:conformally-soft} we derive the conformally soft gluon and graviton theorems in CCFT$_{d-1}$ as a limit of the Ward identities of a shadow current and the stress tensor in CFT$_{d}$. In sections \ref{sec:embedding-gluon}, \ref{sec:embedding-graviton} we revisit the derivation of these Ward identities using the embedding space formalism. The large-$R$ limits of these identities are worked out in section \ref{sec:flat-space}. After projection to the Lorentzian cylinder, we demonstrate in section \ref{sec:gluons} that the leading conformally soft gluon is obtained from the components of the shadow current transverse to the $S^{d-1}$ at $\tau = \frac{\pi}{2}$. The leading and subleading conformally soft gravitons are similarly extracted from an expansion of the transverse traceless component of the shadow stress tensor around $\tau = \frac{\pi}{2}$ in section \ref{sec:gravitons}. We collect various technical results in the appendices.

\section{Preliminaries}\label{sec:preliminaries}

In this section we review how, in the large AdS radius limit, scalar AdS Witten diagrams reduce to Feynman diagram constituents of celestial amplitudes.  This result will be extended to account for massless spinning external states, as well as exchanges of arbitrary mass and spin in section \ref{sec:spin}. Importantly, we clarify the relation between insertions of CFT operators at different global times $\tau_0$ in a strip of width $\Delta \tau = O(R^{-1})$  and the continuum of celestial operators corresponding to an asymptotic state in 4D AFS.

Conformal correlation functions in CFT$_d$ are obtained by summing over all possible AdS$_{d+1}$ Witten diagrams \cite{Witten:1998qj}. The building blocks of the latter are bulk-to-boundary and bulk-to-bulk propagators. It will be convenient to express the bulk-to-boundary propagators in the embedding space formalism \cite{Costa:2011mg,Costa:2014spinning}.  We denote points or vectors in the embedding space $\mathbb{R}^{2,d}$ by capital letters $X$, $P$, $\cdots$. Points in bulk AdS$_{d+1}$ are constrained to obey $X^2 := \eta_{\mu\nu}X^{\mu} X^{\nu} =-R^2$, where $\eta_{\mu\nu} = (-,+,\cdots,+, -)$ and can be parameterized by global coordinates $(\tau,\rho,\vec{z})$ as
\begin{eqnarray}\label{eq:bulk-global-coords}
    X^0(\tau,\rho,\vec{z})&=& R \dfrac{\sin\tau}{\cos\rho},\quad X^{d+1}(\tau,\rho,\vec{z}) = R \dfrac{\cos\tau}{\cos\rho},\quad X^i(\tau,\rho,\vec{z}) = R\tan\rho \Omega^i(\vec{z}).
\end{eqnarray}
Here $\Omega(\vec{z})\in S^{d-1}$ are unit normals to the sphere parameterized by coordinates $\vec{z}$ with
\begin{eqnarray}
\label{eq:sphere-param}
    \Omega(\vec{z}) &=& \left(\dfrac{2z^1}{1+|\vec{z}|^2},\dots,\dfrac{2z^{d-1}}{1+|\vec{z}|^2},\dfrac{1-|\vec{z}|^2}{1+|\vec{z}|^2}\right).
\end{eqnarray}
In these coordinates the boundary is located at $\rho = \frac{\pi}{2}$ and boundary points correspond to null vectors $P^2 = 0$, where
\begin{eqnarray}
    P(\tau,\vec{z}) &=& \lim_{\rho\to\frac{\pi}{2}} \frac{\cos\rho}{R}X(\tau,\rho,\vec{z}),
\end{eqnarray}
or equivalently
\begin{eqnarray}\label{eq:bdry-global-coords}
    P^0(\tau,\vec{z}) &=& \sin\tau,\quad P^{d+1}(\tau,\vec{z})= \cos\tau,\quad P^i(\tau,\vec{z}) = \Omega^i(\vec{z}).
\end{eqnarray}

The correlation functions $\langle {\cal O}_{\Delta_1}(P_1)\cdots{\cal O}_{\Delta_n}(P_n)\rangle$ of scalar operators ${\cal O}_{\Delta_i}(P_i)$ in a holographic CFT$_d$ can be computed by summing over AdS$_{d+1}$ Witten diagrams (see \cite{Penedones:2016voo} for a review). Motivated by the relation between scattering amplitudes and AdS/Witten diagrams in the flat space limit \cite{Penedones:2010ue,Hijano:2019flat,Hijano:2020szl}, a limit was proposed in \cite{PipolodeGioia:2022exe} in which AdS/Witten diagrams reduce to celestial amplitudes.  In this prescription, boundary operators are placed at 
\be 
\label{eq:rescaled-bdry}
\tau_i = \pm\frac{\pi}{2}+\frac{u_i}{R},
\ee 
while bulk global coordinates are redefined as
\be 
\label{eq:bulk-rescaled}
\tau = \frac{t}{R}, \quad \rho = \frac{r}{R},
\ee 
before taking $R\to \infty$ with $(t,r)$ fixed. One of the main observations of \cite{PipolodeGioia:2022exe} is that to leading order at large $R$, scalar bulk to boundary propagators in AdS$_{d+1}$
\begin{eqnarray}
\label{eq:scalar-bulk-to-bdry}
    K_\Delta(X,P)=\dfrac{C_\Delta}{(- P\cdot X+i\epsilon)^\Delta},
\end{eqnarray}
 with $C_{\Delta}$ a normalization constant, become proportional to $\mathbb{R}^{1,d}$ conformal primary wavefunctions \cite{Pasterski:2016qvg} 
\be 
\label{eq:cpw}
\varphi_{\Delta}(x;\eta \hat{q}) = \frac{(i\eta)^{\Delta} \Gamma(\Delta)}{(-\hat{q} \cdot x + i\eta \epsilon)^{\Delta}}.
\ee
 Here $\eta = \pm 1$ depending on whether the boundary operators are placed around $\tau = \pm \frac{\pi}{2}$ with the spheres at $\tau = \pm \frac{\pi}{2}$ assumed to be antipodally related, $x$ is a point in $(d+1)$-dimensional flat space and
\begin{eqnarray}
\label{eq:null-flat}
    \hat{q}(\vec{z}) &=& \left(1, \Omega(\vec{z})\right).
\end{eqnarray}
Analyzing the other elements of the AdS/Witten diagrams, one concludes that these reduce to the building blocks of celestial amplitudes to leading order at large $R$.

The correspondence established in \cite{PipolodeGioia:2022exe} left an important question open. A bulk scalar field in AdS corresponds to an operator of definite dimension in CFT, while massless asymptotic states in flat space should map to a continuum of operators of dimensions $\Delta = \frac{d-1}{2} + i\lambda$ in CCFT$_{d-1}$ \cite{Pasterski:2017kqt}. In contrast, according to \eqref{eq:scalar-bulk-to-bdry}, \eqref{eq:cpw} the celestial amplitudes appear to simply inherit the dimension of the primary operator in the parent CFT.
We conclude this section by explaining how one can in fact extract a continuum of operators in CCFT from the large $R$ expansion of \eqref{eq:scalar-bulk-to-bdry}. 

 Recall that the conformal primary wavefunctions obtained from bulk-to-boundary propagators in the large $R$ limit depend on the position at which the CFT$_d$ operators are inserted within the global time strip of infinitesimal width $\propto R^{-1}$. In particular, 
\be 
\lim_{R\rightarrow \infty} \left. K_{\Delta}(X, P)\right|_{\tau_p = \frac{\pi}{2} - \frac{u_0}{R}} \propto \frac{1}{(t - u_0 - r \Omega \cdot \Omega_p + i\epsilon)^{\Delta}} + O(R^{-1}).
\ee
This result corresponds to an outgoing conformal primary wavefunction defined with respect to a different origin in spacetime, namely
\be 
\varphi_{\Delta}(x - x_0; \hat{q}) \propto \frac{1}{(-\hat{q}\cdot (x - x_0) + i\epsilon)^{\Delta}},
\ee
where $x_0 = (u_0, 0, 0, 0).$
Now note that this shift in origin can be traded for a shift in the conformal dimension $\Delta$ by an integral transform on $u_0$. Specifically,
\be 
\label{eq:Carroll-transform}
\begin{split}
\int_{-\infty}^{\infty} du_0 u_0^{-\Delta_0} \frac{i^{\Delta}}{(t - u_0 - r\Omega \cdot \Omega_p + i\epsilon)^{\Delta}} &= \frac{1}{\Gamma(\Delta)}\int_{-\infty}^{\infty} du_0 u_0^{-\Delta_0} \int_0^{\infty} d\omega \omega^{\Delta - 1} e^{i \omega(t - u_0 - r\Omega \cdot \Omega_p + i\epsilon)}\\
&=  \frac{2 i^{\Delta - 1}\sin (\pi \Delta_0) B(\Delta + \Delta_0 - 1, 1 - \Delta_0)}{(t - r\Omega \cdot \Omega_p + i\epsilon)^{\Delta + \Delta_0 - 1}}, \quad {\rm Re}\Delta_0 \in (0,1),
\end{split}
\ee
where $B(x,y)$ is the Euler beta function.  Similar to calculations involving conformal primary wavefunctions in CCFT, the integral formally converges only for $\Delta_0 = c + i\lambda$, with $c \in (0, 1)$ and $\lambda \in \mathbb{R}$. Nevertheless the result may be analytically continued away from this line in the complex $\Delta_0$ plane \cite{Donnay:2020guq, Freidel:2022skz, Cotler:2023qwh}. Following \cite{Pasterski:2017kqt}, these conformal primary wavefunctions can then be shown to form a complete basis for asymptotic scattering states in $\mathbb{R}^{1,d}$ provided that $\Delta_0$ takes the appropriate continuum of values.

We conclude that up to an interesting normalization,\footnote{In \eqref{eq:Carroll-transform} we assumed that one can exchange the order of integrals over $u_0$ and $\omega$. It would be important, yet beyond the scope of this paper, to study under what conditions this is allowed. It is possible that different prescriptions will yield celestial amplitudes that differ by Poincar\'e invariant structures as observed for example in \cite{Casali:2022fro, Melton:2022fsf}. We thank Walker Melton and Sruthi Narayanan for a discussion on this point. It would also be interesting to understand the precise relation between our prescription and those proposed in \cite{Iacobacci:2022yjo,Sleight:2023ojm} based on an AdS/dS slicing of flat space.} insertions of CFT$_d$ operators at different points in the infinitesimal global time intervals generate the expected continuum of CCFT$_{d - 1}$ operators. The transformation \eqref{eq:Carroll-transform} is the same that maps operators in a Carrollian conformal field theory to celestial operators \cite{Donnay:2022aba, Donnay:2022wvx}. We will return to this in section \ref{sec:ckv-action}. A complementary approach is to keep the $u_0$ dependence and then relate the $R\to \infty$ limit of AdS Witten diagrams to Carrollian correlators instead of celestial ones \cite{Bagchi:2023fbj}.

\section{Spinning celestial amplitudes from flat space limit}
\label{sec:spin}

We now discuss the extension of the result reviewed in the previous section to external spinning operators. We analyze in turn the flat space limit of massless spinning bulk-to-boundary propagators, spinning bulk-to-bulk propagators and vertices.

\subsection{Bulk-to-boundary propagators}

We start by considering the spinning bulk-to-boundary propagators for fields of dimension $\Delta$ and spin $J$ \cite{Costa:2014spinning}
\be 
\label{eq:spinning-b-t-bdry}
K^{\Delta, J}_{\vec{\mu}; \vec{\nu}}(X;P) = C_{\Delta;J}\p_{\mu_1} X^{A_1} \cdots \p_{\mu_J} X^{A_J}\p_{\nu_1} P^{B_1} \cdots \p_{\nu_J} P^{B_J} \frac{I_{\{A_1;\{B_1}(X;P) \cdots I_{A_J\};B_J\}}(X;P)}{(- P\cdot X + i\epsilon)^{\Delta}},
\ee
where 
\be 
\label{I}
I_{A;B}(X;P) = \dfrac{-P\cdot X \eta_{AB}+ P_AX_B}{- P\cdot X+i\epsilon}.
\ee
Here $A_i, B_i$ are $\mathbb{R}^{2,d}$ embedding space indices, $\mu_i$ run over the rescaled coordinates $(t, r, \Omega)$ defined in \eqref{eq:sphere-param}, \eqref{eq:bulk-rescaled} and $\nu_i$ run over the boundary coordinates $(u, \Omega)$ in \eqref{eq:rescaled-bdry}. $\p_{\mu_i} X^{A_i}$, $\p_{\nu_i} P^{B_i}$ hence implement projections onto the corresponding bulk and boundary tensors respectively and $\{ \cdot \}$ denotes the symmetric traceless component. We collect some useful results on the embedding space formalism in appendix \ref{sec:embedding-space}. 
$C_{\Delta,J}$ is a normalization constant \cite{Costa:2014spinning} 
\begin{eqnarray}
    C_{\Delta,J} &=&\dfrac{(J+\Delta-1)\Gamma(\Delta)}{2 \pi^{d/2}(\Delta-1)\Gamma(\Delta+1-\frac{d}{2}){R^{(d-1)/2-\Delta+J}}}.
\end{eqnarray}
We see that spinning bulk-to-boundary propagators are obtained from the scalar ones defined in \eqref{eq:scalar-bulk-to-bdry} by dressing with the conformally covariant tensors in \eqref{I}. It then suffices to analyze the behavior of these tensors in the flat space limit. 

Using the large $R$ expansions 
\begin{eqnarray}
    X(\tau,\rho,\vec{z}) &=& (0,R) + (x,0) + O(R^{-1}),\label{eq:expansion-bulk-embedding}\\
    P(\tau_i,\vec{z}_i) &=& \pm (\hat{q}(\vec{z}_i),0) \mp \left(0,\frac{u_i}{R}\right)+O(R^{-2})\label{eq:expansion-bdry-embedding}
\end{eqnarray}
 of the bulk and boundary embedding space vectors,
where $x = (t,r\Omega(\vec{z}))$ are Cartesian coordinates and $\hat{q}$ is defined in \eqref{eq:null-flat},
 one obtains the expansions of the projectors $\partial_\mu X^A$ and $\partial_\nu P^B$.
From these expansions it immediately follows that
\begin{eqnarray}
    \eta_{AB}\partial_\mu X^A\partial_\nu P^B&=& \begin{cases}O(R^{-2}),& \nu =u,\\ \pm \partial_a \hat{q}_{\mu}(\vec{z}) +O(R^{-1}),& \nu=z^a, \end{cases}\label{eq:projector-id-1}\\
    P_A X_B\partial_\mu X^A\partial_\nu P^B &=& \begin{cases} \hat{q}_\mu(\vec{z})+O(R^{-1}),& \nu=u,\\ \left(\partial_a \hat{q}(\vec{z})\cdot x\right) \hat{q}_\mu(\vec{z})+O(R^{-1}),& \nu=z^a.\end{cases}\label{eq:projector-id-2}
\end{eqnarray}

The expansion of the conformally covariant tensors \eqref{I} projected onto bulk and boundary indices follows directly from these results. We  distinguish between two cases. First, when the boundary index is $\nu = u$ we have
\begin{eqnarray}
I_{\mu,u}(X,P) &=& \pm\lim_{\Delta \rightarrow 0} \frac{1}{\Delta}\left[\partial_\mu\left(\dfrac{1}{(- \hat{q}\cdot x\pm i\epsilon)^{\Delta}}\right)+O(R^{-1})\right]\label{eq:flat-limit-bulk-bdry-1},
\end{eqnarray}
which we recognize as the derivative of a scalar conformal primary wavefunction. Likewise, if the boundary index is $\nu = z^a$ we have
\begin{eqnarray}
I_{\mu, a}(X,P) 
    &=& \pm \left[\partial_a \hat{q}_\mu(\vec{z}) +\dfrac{\partial_a \hat{q}(\vec{z})\cdot x }{(- \hat{q}\cdot x\pm i\epsilon)}\hat{q}_\mu(\vec{z})+O(R^{-1})\right].
    \label{eq:flat-limit-bulk-bdry-2}
\end{eqnarray}
Hence, up to normalization and a phase, the flat space limit of $I_{\mu, a}(X,P)$ corresponds to the conformally covariant tensor used in the construction of spinning conformal primary wavefunctions given in \cite{Pasterski:2017ylz}.\footnote{The polarization vectors $\p_a \hat{q}$ are gauge equivalent to the ones defined in \cite{Pasterski:2017kqt}.}
 Putting everything together, we conclude that general massless spinning conformal primary wavefunctions are obtained from  flat space limits of the spinning bulk-to-boundary propagators \eqref{eq:spinning-b-t-bdry} with transverse indices. Note however that the dimensionally reduced bulk to boundary propagators have a non-vanishing trace. In order to obtain conformal primary wavefunctions in CCFT$_{d-1}$ the trace has to be subtracted. For example, in the spin two case this is implemented by applying the projector \cite{Pasterski:2017kqt}
 \be 
 \label{eq:projector}
 P^{b_1b_2}_{a_1 a_2} \equiv \delta^{b_1}_{\{a_1} \delta^{b_2}_{a_2\}} - \frac{1}{d - 1} \delta_{a_1 a_2} \delta^{b_1b_2}.
 \ee

Finally, \eqref{eq:flat-limit-bulk-bdry-1} implies that bulk-to-boundary propagators with time indices on the boundary result in pure gauge conformal primary wavefunctions. We leave a better understanding of this, as well as additional data resulting from the dimensional reduction to future work.

\subsection{Bulk-to-bulk propagators and vertices}

The spin $J$ bulk-to-bulk propagator in AdS$_{d+1}$ obeys the equations \cite{Costa:2014spinning}
\begin{eqnarray}
    \left(\Box_{AdS}-\frac{\Delta(\Delta-d)}{R^2}+\frac{J}{R^{2}}\right)\Pi_{\mu_1\dots \mu_J,\nu_1\dots \nu_J}(X,\bar X) &=& -g_{\mu_1\{\nu_1}\cdots g_{|\mu_J|\nu_J\}}\delta_{AdS}(X,\bar X),\\
    \nabla^{\mu_1}\Pi_{\mu_1\dots \mu_J,\nu_1\dots \nu_J}(X,\bar X) &=&0.
\end{eqnarray}
To take the flat space limit we assume that all of the components are in the chart $(t,r,\Omega)$, in which the AdS metric $g_{\mu\nu}$ becomes the Minkowski metric $\eta_{\mu\nu}$ to leading order at large $R$ 
\begin{eqnarray}
    g_{\mu\nu} = \eta_{\mu\nu}+O(R^{-2}).
\end{eqnarray}
On the other hand, the Laplace operator behaves as $\Box_{AdS}=\Box_{\mathbb{R}^{1,d}}+O(R^{-2})$ and the Dirac delta behaves as $\delta_{AdS}(X,\bar X)=\delta_{\mathbb{R}^{1,d}}(x,\bar x)+O(R^{-2})$ \cite{PipolodeGioia:2022exe}. Therefore the first equation turns into the equation for the propagator of a spin $J$ field of mass $m=\underset{R\to \infty}{\lim}\frac{\Delta}{R}$ in flat space. The second equation can be treated in the same way since $g_{\mu\nu} = \eta_{\mu\nu}+O(R^{-2})$ and the AdS covariant derivative becomes the flat spacetime covariant derivative when $R\to \infty$.

As a result, the bulk-to-bulk propagator must have an expansion of the form
\begin{eqnarray}
    \Pi_{\mu_1\dots \mu_J,\nu_1\dots \nu_J}(X,\bar X) &=& G_{\mu_1\dots \mu_J,\nu_1\dots \nu_J}(x,\bar x)+O(R^{-2}),
\end{eqnarray}
where $G_{\mu_1\dots \mu_J,\nu_1\dots \nu_J}(x_1,x_2)$ is the Feynman propagator for a symmetric traceless tensor of spin $J$ in $\mathbb{R}^{1,d}$. 

Since vertices are simply integrals over AdS which become integrals over $\mathbb{R}^{1,d}$ in the flat space limit, we conclude that AdS-Witten diagrams for spinning particles reduce to CCFT$_{d-1}$ amplitudes of spinning massless particles in the flat space configuration \eqref{eq:rescaled-bdry}.

\section{From conformal to infinite dimensional symmetry}
\label{sec:infinite-dimensional}

Consider a $d$-dimensional CFT on the Lorentzian cylinder with metric
\be 
\label{eq:cyl-metric}
ds^2 = g_{\mu\nu} dx^{\mu} dx^{\nu} = -d\tau^2 + d\Omega^2_{d - 1},
\ee
where $d\Omega^2_{d - 1}$ is the metric on the $(d-1)$-sphere of unit radius. Conformal transformations are coordinate transformations that preserve the metric up to a  Weyl rescaling. Specifically, infinitesimal conformal transformations are obtained by finding the diffeomorphisms
\be 
x^{'\mu} = x^{\mu} + \epsilon^{\mu}(x)
\ee
under which the metric transforms as
\be 
g_{\mu\nu}'(x') = g_{\mu\nu}(x) + \delta g_{\mu\nu}, \quad \delta g_{\mu\nu} = \sigma(x) g_{\mu\nu}(x).
\ee
Such diffemorphisms are subject to the conformal Killing equations
\be 
\label{eq:CK}
\nabla_{\mu} \epsilon_{\nu} + \nabla_{\nu} \epsilon_{\mu} = \frac{2}{d} \nabla \cdot \epsilon(x) g_{\mu\nu}.
\ee
The solutions to these equations generate the conformal algebra $\mathfrak{so}(d, 2)$ for $d \geq 3$, while for $d = 2$ this algebra admits a Virasoro enhancement.  

The relation between celestial amplitudes on the $(d-1)$-dimensional celestial sphere and conformal correlation functions of primary operators localized to strips of infinitesimal width $\Delta \tau \propto \frac{1}{R}$ as $R \rightarrow \infty$ suggests that,  on short global time scales, $d$-dimensional conformal field theories should develop an infinite dimensional symmetry. In this section we show that this is indeed the case by analyzing the conformal Killing equations \eqref{eq:CK} in this limit. We specialize to $d = 3$ in which case the emergent ``celestial'' CFT is 2-dimensional and expected to be governed by the extended BMS symmetries of 4D asymptotically flat spacetimes  (AFS) \cite{Barnich:2011ct, as1, He:2014laa, Kapec:2014opa}. 

For $d = 3$, \eqref{eq:cyl-metric} reduces to
\be 
\label{eq:3D-cyl}
ds^2 = -d\tau^2 + 2\gamma_{z\bz} dz d\bz, \quad \gamma_{z\bz} = \dfrac{2}{(1 + z\bz)^2},
\ee
where we introduced stereographic coordinates $(z, \bz)$ on the unit 2-sphere with metric $\gamma_{z\bz}$. 
We would like to zoom into a region of the 3-dimensional Lorentzian cylinder of infinitesimal width centered around a global time slice at $\tau_0$. To this end, we introduce the coordinate $u$ defined by
\be 
\label{tau}
\tau = \tau_0 + \frac{u}{R},
\ee
in which case the metric \eqref{eq:3D-cyl} becomes
\be 
\label{metric-exp}
ds^2 =  -R^{-2} du^2 +  2\gamma_{z\bz} dz d\bz.
\ee
The conformal Killing equations associated with \eqref{metric-exp}  take the form
\begin{eqnarray}
\label{ckv}
\p_u \epsilon^u &= &  \frac{1}{3}\nabla \cdot \epsilon, \label{1} \\
\p_u \epsilon_z + \p_z \epsilon_u &=& 0, \label{2} \\
D_{\bz} \epsilon_z + D_z \epsilon_{\bz} &=& \frac{2}{3} \nabla \cdot \epsilon \gamma_{z\bz}, \quad D_z \epsilon_z = 0, \label{3}
\end{eqnarray}
where $D_A$ is the covariant derivative on the sphere and we denote indices tangent to the sphere by $A$. 

The last equation in \eqref{3}  is solved by
 \be\label{ea} 
\begin{split}
\gamma_{z\bz} \p_z \epsilon^{\bz} &= \gamma_{z\bz} \p_{\bz} \epsilon^z = 0 \implies \epsilon^A = F(u) Y^A(z,\bz),
\end{split}
\ee
where $Y^A$ are conformal Killing vectors on the sphere. Moreover \eqref{1} and the first equation in \eqref{3} yield\footnote{Note that $f(z,\bz)$ may depend on $R$. As we show later, the global translations are obtained from an Inonu-Wigner contraction of vector fields with $f(z,\bz) = R$. Supertranslations may also be obtained by allowing $f(z,\bz) = R f_0(z,\bz)$ and directly applying \eqref{Wigner-momentum} to the local generators.}
\be 
\label{eu}
2 \p_u \epsilon^u = F(u) D \cdot Y  \implies \epsilon^u = \frac{1}{2}\int^u du' F(u') D\cdot Y + f(z, \bz).
\ee
Finally, $F(u)$ is determined from \eqref{eu} and \eqref{2}. In the limit as $R \rightarrow \infty$ we distinguish between two cases. If $D\cdot Y = 0$ we immediately find
\be 
\label{eq:partial-sol-rot}
\p_u F(u) = O(R^{-2}) \implies F(u) = c + O(R^{-2}),
\ee 
where $c$ is a constant. For future convenience we chose $c = 1$ which reproduces the standard Lie algebra of rotation generators to leading order at large $R$. On the other hand, if $D\cdot Y \neq 0$, taking a $u$ derivative of \eqref{2} we find
\be 
\p_u^2 F(u) Y_A - \frac{F(u)\p_A D\cdot Y}{2 R^2} = 0,
\ee
or upon taking the divergence on the sphere,\footnote{Recall that conformal Killing vectors  on the sphere obey \be 
D_AD^A D_B Y^B = -2 D\cdot Y.
\ee}
\be 
\label{F-diff-eq}
\left[\p_u^2 F(u) + \frac{1}{R^2} F(u)\right] D\cdot Y = 0.
\ee
\eqref{F-diff-eq} is solved by 
\be 
\label{eq:partial-sol-boost}
F(u) = e^{\pm i(\tau_0 + \frac{u}{R})}.
\ee
Since we have taken a $u$ derivative and a divergence on the sphere in order to arrive at \eqref{eq:partial-sol-rot} and \eqref{eq:partial-sol-boost}, it is important to verify whether these solutions also obey the original conformal Killing equation \eqref{2}. In fact \eqref{eq:partial-sol-rot}, \eqref{eq:partial-sol-boost} fail to obey \eqref{2} away from the $R \rightarrow \infty$ limit.  For $D \cdot Y \neq 0$
\be 
\label{uz-constr}
\delta_{\epsilon^{\pm}} g_{uA} = \pm \frac{i e^{\pm i(\tau_0 + \frac{u}{R})}}{R}  \alpha_A(z,\bz) - \frac{\p_A  f(z,\bz)}{R^2}, \quad \quad \alpha_A = Y_A+\dfrac{1}{2}D_A(D\cdot Y).
\ee
Therefore the violation is $O(R^{-1})$ for the local CKV on the sphere, while in the special case $D \cdot Y = 0$ the violation is $O(R^{-2})$. The enhanced conformal Killing symmetry in the strip is therefore broken at $O(R^{-1})$. Singularities in the local CKVs on the sphere also lead to a violation of the conformal Killing equations by contact terms.

The vector fields that preserve the metric of a 3D Lorentzian cylinder in an infinitesimal time interval $\propto R^{-1}$ in the limit $R \rightarrow \infty$ are hence
\be 
\label{asy-ckv}
\epsilon^{\pm} = \left[\mp \frac{iR}{2} F_{\pm}(u) D\cdot Y + f(z, \bz)\right] \p_u + F_{\pm}(u) Y^A \p_{A},
\ee
 where 
\be 
\label{F-sol-fin}
\begin{cases}
F_{\pm}(u) = e^{\pm i(\tau_0 + \frac{u}{R})}, \quad D\cdot Y \neq 0,\\
F_{\pm}(u) = 1, \quad D\cdot Y = 0.
\end{cases}
\ee
 It may be interesting, yet beyond the scope of this paper, to systematically understand whether \eqref{metric-exp} and \eqref{asy-ckv} admit subleading corrections\footnote{Unfortunately this naively appears to require coupling the boundary CFT to gravity. We thank Jan de Boer for a discussion on this point.} at large $R$ that allow for an enhancement of conformal symmetry in a strip of small yet finite size.

A few comments are in order. Just like the generators of the extended BMS group in 4D AFS, the vector fields \eqref{asy-ckv} are labelled by a function $f(z, \bz)$ and a local conformal Killing vector $Y^A(z,\bz)$ on the sphere. The resulting symmetry group is infinite dimensional, in contrast to the conformal group in 3 dimensions. At first glance this may seem surprising, however we ought to keep in mind that \eqref{asy-ckv} are \textit{not} symmetries of full 3D CFT but only of infinitesimal time intervals. On the other hand, global conformal symmetry relates short and long time scales so it may be possible to reinterpret the integral transform in \eqref{eq:Carroll-transform} as an integral over the whole cylinder.\footnote{We thank Laurent Freidel for a discussion on this point.} 

Moreover, note that in the $R \rightarrow \infty$ limit the metric \eqref{metric-exp} develops a ``null direction'' reflected by the vanishing of the $g_{uu}$ component. As such, the restriction to short global timescales shares similarities with the Carrollian limit \cite{Levy1965, Bacry:1968zf}. In the next section we show how the extended BMS$_4$ algebra is recovered from the enhanced conformal symmetries \eqref{asy-ckv} of the strip by a Inonu-Wigner contraction \cite{Inonu}.

\subsection{Extended BMS$_4$ algebra in CFT$_3$}
\label{subsec:extendedBMS}

We now show that the extended BMS$_4$ algebra can be extracted from the algebra generated by the vector fields \eqref{asy-ckv}. This procedure is analogous to Inonu-Wigner contraction of the conformal algebra to Poincar\'e \cite{Inonu}. 

We start by noting that appropriate linear combinations of \eqref{asy-ckv} generate an $\mathfrak{ so}(3,2)$ algebra for constant $f(z, \bz)$ and $Y = Y^A\p_A$ restricted to the global conformal Killing vectors of the sphere \cite{Kapec:2014opa},
\begin{equation}
\label{eq:rot-sph}
\begin{split}
    Y_{12} &= -i(z\partial_z-\bar z \partial_{\bz}),\quad
    Y_{23} = -i\dfrac{z^2-1}{2}\partial_z+i\dfrac{\bar z^2-1}{2} \partial_{\bz},\quad
    Y_{31} = -\dfrac{1+z^2}{2}\partial_z-\dfrac{1+\bar z^2}{2}\partial_{\bz},
    \end{split}
\end{equation}
\begin{equation}
\label{eq:boosts}
\begin{split}
    Y_{01} &= \dfrac{1-z^2}{2}\partial_z +\dfrac{1-\bar z^2}{2}\partial_{\bz},\quad
    Y_{02} = \dfrac{i(1+z^2)}{2}\partial_z-\dfrac{i(1+\bar z^2)}{2}\partial_{\bz},\quad
    Y_{03} = -z\partial_z -\bar z \partial_{\bz}.
    \end{split}
\end{equation}
\eqref{eq:rot-sph} correspond to rotations of the 2-sphere and have vanishing divergence $D\cdot Y_{ij} = 0$ while \eqref{eq:boosts} have non-vanishing divergence
\be 
D\cdot Y_{0i} = -2 \Omega_i,
\ee
where $\Omega = \frac{1}{1 + z\bz}\left(z + \bz, -i(z - \bz), 1 - z\bz \right)$ is the unit normal to the sphere at $(z,\bz).$ Specifically, identifying 
\begin{eqnarray}
	D &=& {-i}\epsilon_{f=R},\quad
	J_{ij} = {i}\epsilon_{Y_{ij}},\\
	P_i &=& {i}\epsilon^+_{Y_{0i}} ,\quad~~
	K_i = {i}\epsilon^-_{Y_{0i}} ,
\end{eqnarray}
we find the commutation relations \cite{Penedones:2016voo} 
\be 
\label{eq:conf-algebra}
\begin{split}
    [D, J_{ij}] &= 0, \quad [D, P_i] = P_i, \quad [D, K_i] = - K_i, \\
    [J_{ij}, P_k] &= i(\delta_{ik}P_j - \delta_{jk}P_i),\quad [J_{ij}, K_k] = i(\delta_{ik}K_j - \delta_{jk}K_i),\\
    [P_i, K_j] &= 2i(i \delta_{ij}D  -J_{ij}), \quad {[J_{ij}, J_{k\ell}] = i \left[ \delta_{ik} J_{j \ell} + \delta_{j\ell} J_{ik} -\delta_{jk} J_{i\ell} - \delta_{i\ell} J_{jk} \right]}.
\end{split}
\ee

These generators can be reorganized in terms of Lorentz generators $M_{AB}$ of the embedding space $\mathbb{R}^{2,3}$ \footnote{{Our conventions differ slightly from those in \cite{Penedones:2010ue} and are simply related by exchanging the $0$ and $4$ directions or equivalently shifting $\tau \rightarrow \tau + \frac{\pi}{2}$ in \eqref{eq:bulk-global-coords}}.}
\begin{eqnarray}
    M_{40} &=& -D,\quad M_{i4} = \dfrac{P_{i} + K_i}{2},\\
    M_{ij} &=& J_{ij},\quad M_{i0} =\dfrac{P_i - K_i}{2i}, \quad i = 1,2,3.
\end{eqnarray}
Explicit computation shows that \eqref{eq:conf-algebra} imply that $M_{AB}$  obey the $\mathfrak{ so}(3,2)$ algebra
\be 
\label{Lorentz-algebra}
{[M_{AB},M_{CD}]} =i(\eta_{AC}M_{BD}+\eta_{BD}M_{AC}-\eta_{BC}M_{AD}-\eta_{AD}M_{BC})
\ee
with $\eta_{00} = \eta_{44} = -1, \eta_{ii} = 1$ and all other components vanishing. 
The Inonu-Wigner contraction is implemented by redefining 
\be 
\label{Wigner-momentum}
  {\cal P}^{\mu} = \dfrac{1}{R}M^{4\mu}, \quad \mu = 0, \cdots, 3
\ee
and taking $R \rightarrow \infty$ while keeping $\mathcal{P}^{\mu}$ and $M_{\mu\nu}$ fixed. It is straightforward to show that in this limit, \eqref{eq:conf-algebra} reduce to the Poincar\'e algebra, with $\mathcal{P}^{\mu}$ and $M_{\mu\nu}$ the translation and Lorentz generators in $\mathbb{R}^{1,3}$ respectively. 

\begin{figure}[h!]
\begin{center}
\begin{tikzcd}[row sep=large, column sep=80pt]
\mathfrak{so}(3,2)~ {\rm generators} \arrow{r}{{\rm short~times}} \arrow{d}[swap]{{\rm Inonu-Wigner}} & {\rm local~enhancement ~\eqref{asy-ckv}} \arrow{d}{{\rm Inonu-Wigner}} \\   
{ \text{Poincar\'e}} \arrow[swap]{r}{{}} & \mathfrak{ebms}_4
\end{tikzcd}
\caption{The metric of a CFT$_d$ on the Lorentzian cylinder develops an approximately null direction over infinitesimal global time intervals $\Delta \tau \sim R^{-1}$. In the limit $R \rightarrow \infty$, the conformal Killing equations admit an infinite dimensional set of solutions parameterized by a function on $S^{d-1}$ and a conformal Killing vector on $S^{d-1}$. In particular, for $d = 3$, an Inonu-Wigner contraction in the intervals around $\tau = \pm \frac{\pi}{2}$ leads to vector fields that obey the extended BMS$_4$ algebra.}
\label{fig:contraction}
\end{center}
\end{figure}
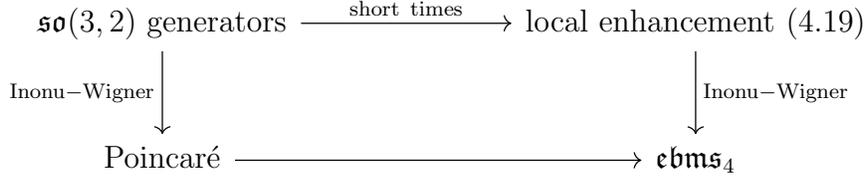

We now demonstrate that an analogous Inonu-Wigner contraction of the local vector fields \eqref{asy-ckv} leads to the extended BMS$_4$ algebra $\mathfrak{ebms}_4$. In analogy with \eqref{Wigner-momentum} we define 
\begin{eqnarray}
\label{eq:TL}
    T_Y &=& i\dfrac{\epsilon_{Y}^++\epsilon_Y^-}{2R}, \quad L_Y = \dfrac{\epsilon^+_Y-\epsilon^-_Y}{2}
\end{eqnarray}
for arbitrary conformal Killing vector fields $Y$\footnote{Note that the rotation generators with $D\cdot Y = 0$ are obtained directly as $M_{ij} = J_{ij}$, hence no linear combination is necessary.} and take the limit $R \rightarrow \infty$. Setting $\tau_0 = \frac{\pi}{2} + O(R^{-1})$, we find from \eqref{asy-ckv} and \eqref{eq:TL}
\begin{eqnarray}
    -i T_Y &=& \dfrac{1}{2}D\cdot Y\partial_u+O(R^{-2}),\\
      -i L_Y &=& Y^A\partial_A + \frac{u}{2}D\cdot Y \partial_u +O(R^{-2})\label{eq:lorentz-generators} \label{eq:super-rot}.
\end{eqnarray}
Together with the vector fields with $Y= 0$, parametrized by an arbitrary function $f$ on the sphere
\be
T_f \equiv i\epsilon_f = i f(z,\bar z)\partial_u+O(R^{-2}),
\ee
$L_Y$ generate $\mathfrak{ebms}_4$ 
\begin{eqnarray}
    [T_{f_1},T_{f_2}] &=&O(R^{-2}),\\ {[}L_{Y_1},L_{Y_2}]&=& i L_{[Y_1,Y_2]}+O(R^{-2}),\\
    {[}T_{f},L_{Y}] &=& \left[Y(f) - \dfrac{1}{2}(D\cdot Y)f(z,\bar z)\right]\partial_u+O(R^{-2}) = i T_{f' = \frac{1}{2}(D\cdot Y) f - Y(f)} + O(R^{-2}).
\end{eqnarray}
Note that 
\begin{eqnarray}
   \lim_{R\to \infty}T_Y = \lim_{R\to \infty} T_{f=\frac{1}{2}D\cdot Y}
\end{eqnarray}
which means that $T_Y$ correspond to a special class of  supertranslation vector fields $T_f$ with $f = \frac{1}{2}D\cdot Y$ and are hence redundant. Analogous results are obtained by expanding \eqref{eq:TL} around $\tau_0 = -\frac{\pi}{2}.$ The results of this section are summarized in Figure \ref{fig:contraction}. 

Finally, consider the shift $\tau_0\to \tau_0+\pi$ in $\epsilon_Y^{\pm}$ defined in \eqref{asy-ckv}. Under this transformation, $\epsilon_Y^\pm\to -\epsilon_Y^{\pm}$. The same transformation can be implemented for the globally defined vector fields by keeping $\tau$ fixed and considering instead an antipodal map on $S^2$. Therefore, the action of $L_Y$ and $T_Y$ on $S^2$ slices of the Lorentzian cylinder separated by $\pi$ in global time becomes the same provided the slices are antipodally related. This is compatible with the observation in \cite{PipolodeGioia:2022exe} that in order to respect Lorentz invariance in the flat space limit of AdS Witten diagrams it is necessary to antipodally identify the time-slices corresponding to in/out states. It further suggests that the antipodal matching condition between ${\cal I}^+_-$ and ${\cal I}^-_+$ employed in AFS \cite{as1} arises naturally in the flat space limit proposed in \cite{PipolodeGioia:2022exe}. Note that similar arguments led to a derivation of the matching conditions via a resolution of $i^0$ with hyperbolic slices \cite{Prabhu:2021cgk,Capone:2022gme}. %

\subsection{Transformation of CFT$_3$ primary operators in the strip }\label{sec:ckv-action}

We now study the action of the conformal Killing vectors on CFT$_3$ primary 
operators and show that when restricted to global time slices, these operators transform as quasi-primary operators in CCFT$_2$. We work in Euclidean signature and Wick rotate at the end. 

A primary operator ${\cal O}_\Delta(x)$ of arbitrary spin transforms in some representation $D:{\rm SO}(3)\to {\rm GL}(V)$. The action action of a conformal Killing vector $\epsilon$ on such an operator is \cite{Osborn2019:lectures}
\begin{eqnarray}\label{eq:3d-ckv-action}
    \delta_\epsilon{\cal O}_\Delta(x) &=& -\left[(\nabla\cdot\epsilon)\dfrac{\Delta}{3}+\epsilon^\mu \nabla_{\mu}+\dfrac{i}{2}\nabla_\mu\epsilon_\nu S^{\mu\nu}\right]{\cal O}_\Delta(x),
\end{eqnarray}
where $\nabla_\mu$ is the spin covariant derivative \cite{He:2021covariant} \footnote{This agrees with the definition involving $\Sigma$ in \cite{He:2021covariant} upon setting $\Sigma^{\mu\nu} = iS^{\mu\nu}$, with $S^{\mu\nu}$ obeying \eqref{Lorentz-algebra}. }
\begin{eqnarray}
\label{eq:spin-cov-der}
    \nabla_\mu = \partial_\mu +\frac{i}{2}\omega_\mu^{\phantom \mu ab}S_{ab}.
\end{eqnarray}
Here $\omega_\mu^{\phantom \mu ab}$ is the torsion-free spin connection defined in terms of a vielbein $e_\mu^a$ 
\be 
\label{metric}
g_{\mu\nu} = e_{\mu}^a e_{\nu}^b \delta_{ab},
\ee
where $g_{\mu\nu}$ is the 3-dimensional metric, $S_{ab}$ are the generators of the representation $D$ and $S_{\mu\nu}=e_\mu^a e_\nu^b S_{ab}$.
Note that ${\cal O}_\Delta(x)$ are defined to only carry internal indices. As an example, in appendix \ref{example} we demonstrate that \eqref{eq:spin-cov-der} reduces to the standard Levi-Civita connection when acting on Lorentz vectors. 
The (Wick rotated) metric \eqref{metric-exp} is recovered with the following choice of vielbein $e_\mu^a$ 
\begin{eqnarray}
    e^1 = \sqrt{\dfrac{\gamma_{z\bar z}}{2}}(dz+d\bar z),\quad e^2 =-i\sqrt{\dfrac{\gamma_{z\bar z}}{2}}(dz-d\bar z),\quad e^3 = \dfrac{du}{R}.
\end{eqnarray}

Taking $\epsilon = L_Y$, namely 
\begin{eqnarray}
    L_Y &\equiv& \dfrac{\epsilon_Y^+-\epsilon_Y^-}{2}\label{eq:defn-epsilon_Y}\\
    &=&\frac{i}{2}(D\cdot Y) u\partial_u + iY^A\partial_A+O(R^{-1}), \quad \tau_0 = \frac{\pi}{2},
\end{eqnarray}
we show in appendix \ref{app:ckv-action} that \eqref{eq:3d-ckv-action} becomes 
\begin{equation}\label{eq:ckv-action-flat-limit-1}
    \delta_{L_Y}{\cal O}_\Delta(x) = - i \left[D_zY^z \mathfrak{h} +D_{\bar z}Y^{\bar z}\bar{\mathfrak{ h}}+Y^z(\partial_z-\Omega_z J_3)+Y^{\bar z}(\partial_{\bar z}-\Omega_{\bar z}J_3)+O(R^{-1})\right]{\cal O}_{\Delta}(x),    
\end{equation}
where we defined the operator-valued weights
\begin{eqnarray}
    \mathfrak{h} \equiv \dfrac{\hat{\Delta}+J_3}{2},\quad \bar{ \mathfrak{h}} \equiv\dfrac{\hat{\Delta}-J_3}{2},\quad \hat{\Delta} \equiv  \Delta+u\partial_u.
\end{eqnarray}
Finally given that $J_3$ acts diagonally on a primary operator, 
\begin{eqnarray}
    J_3{\cal O}_\Delta = s {\cal O}_\Delta,
\end{eqnarray}
 the operator-valued weights simplify to
\begin{eqnarray}
\label{eq:weights}
    \mathfrak{h} = \dfrac{\hat{\Delta}+s}{2},\quad \bar{\mathfrak{h}} = \dfrac{\hat{\Delta}-s}{2}.
\end{eqnarray}

On the other hand, note that the dilatation operator in the two-dimensional theory is not diagonal in the basis of primary operators of the CFT$_3$. Indeed, only operators placed at $u=0$ diagonalize the two-dimensional weights \eqref{eq:weights}. For this special case, one obtains operators transforming like two-dimensional primary operators with respect to conformal transformations of the slices, whose dimensions agree with those of the corresponding CFT$_3$ operators. More generally $\hat{\Delta}$ can be diagonalized by the time Mellin-like transform discussed at the level of the bulk-to-boundary propagators in section \ref{sec:preliminaries}, namely
\begin{eqnarray}
\label{eq:time-Mellin}
    \widehat{\cal O}_\Delta(z,\bar z;\Delta_0)\equiv N(\Delta,\Delta_0)\int_{-\infty}^{\infty}du\ u^{-\Delta_0}{\cal O}_\Delta(u,z,\bar z),
\end{eqnarray}
where $N(\Delta, \Delta_0)$ is chosen to reproduce the standard normalization of CCFT operators. 
Under this transformation we have
\begin{eqnarray}
    u\partial_u  ~\rightarrow~ \Delta_0 -1
\end{eqnarray}
and therefore it follows that $\widehat{\cal O}_{\Delta}(z,\bar z;\Delta_0)$ transforms as a two-dimensional quasi-primary operator with weights
\begin{eqnarray}
\label{eq:effective-dim}
    h &=& \dfrac{(\Delta+\Delta_0-1)+s}{2},\quad \bar h = \dfrac{(\Delta+\Delta_0-1)-s}{2}.
\end{eqnarray}
The transformation of $\widehat{{\cal O}}_\Delta$ under $L_Y$ is therefore
\begin{equation}\label{eq:2d-conf-transf-primary-operator-from-3d}
    \delta_{L_Y}\widehat{{\cal O}}_\Delta(z,\bz;\Delta_0) = -i\left[D_zY^z h +D_{\bar z}Y^{\bar z}\bar{h}+Y^z(\partial_z-s\Omega_z)+Y^{\bar z}(\partial_{\bar z}-s\Omega_{\bar z})+O(R^{-1})\right]{\cal \widehat{O}}_{\Delta}.
\end{equation}

As an example consider a  CFT$_3$ current $J_\mu$ of dimension $\Delta = 2$ and spin $s=1$. According to \eqref{eq:2d-conf-transf-primary-operator-from-3d} its restriction to an equal time slice, $(\widehat{J}_z, \widehat{J}_{\bz})$, transforms under 2d conformal transformations of the slice as an operator of dimension $\Delta_{CCFT} = 1 +\Delta_0$ and spin $s=1$. Choosing $\Delta_0=0$ then yields a 2D current. Likewise the stress tensor $T_{\mu\nu}$ has $\Delta_{CFT}=3$ and spin $s=2$. In this case its 2D counterpart $\widehat{T}$ has $\Delta_{CCFT}= 2 +\Delta_0$. Therefore choosing $\Delta_0=0$ again yields an operator that transforms as the stress tensor in two-dimensions. Currents in the dimensionally reduced theory can be equivalently obtained from currents in the parent CFT$_3$ by performing a 3D shadow transform followed by restriction to the $u = 0$ slice and a 2D shadow transform. It can be easily checked that this prescription lowers the dimension of the operator by 1. This is detailed in appendix \ref{app:shadow-dim-red} and motivates our calculations in the following section.

This discussion brings the proposed projection from CFT$_d$ to CCFT$_{d-1}$ closer to the standard dimensional reduction procedure. The starting point in dimensional is a manifold $M\times K$, where $K$ is usually taken to be compact. A field $\Phi$ in this higher-dimensional space can be decomposed into modes that diagonalize a differential operator on $K$. The coefficients in the expansion of $\Phi$ in terms of these modes are then a tower of fields $\Phi_m$ in $M$ \cite{Freedman:2012supergravity}. This is analogous to what happens here. Explicitly, we start with a CFT$_3$ on $\mathbb{R}\times S^2$ and note that the operator ${\cal O}_\Delta(u,z,\bar z)$ can be expanded in terms of eigenfunctions of the differential operator $u\partial_u$ in $\mathbb{R}$ and a continuum of modes $\widehat{\cal O}_\Delta(z,\bar z;\Delta_0)$. In this case the role of $K$ is played by the non-compact $\mathbb{R}$ and therefore we obtain a continuum instead of a discrete set of fields in the dimensionally-reduced theory on $S^2$. Similar ideas applied to the distinct context of relating celestial holography to holography for the continuum of AdS$_3$/CFT$_2$ slices of the future/past Milne wedges of Minkowski spacetime have been put forward in \cite{deBoer:2003vf, Cheung:2016iub}. It would be interesting to establish a precise equivalence between these two approaches. 

Finally, note that the transformation \eqref{eq:time-Mellin} is the same as the one recently employed in \cite{Donnay:2022aba,Donnay:2022wvx} to relate Carrollian and celestial holography. This transformation appears here in a novel context  and we believe it deserves further study. One difference here is that the effective dimension of the CCFT operator is not simply $\Delta_0$, but instead $\Delta + \Delta_0- 1 $. One hence has to account for the shift by the dimension $\Delta$ of the operator in the parent CFT$_3$ when taking conformally soft limits for example. The additional shift by $1$ is due to the fact the CFT$_3$ vector field \eqref{eq:super-rot} has no radial component. In the case of superrotation vector fields in AFS this is known to induce a shift by 1 in the conformal primary dimension of on asymptotic field with respect to its action \cite{Kapec:2014opa}. It would be interesting to further explore how radial evolution  in AFS arises from the perspective of the flat space limit of CFT$_3$.

We conclude this section by noting that in the case when $Y$ is a globally defined CKV on $S^2$, the vector fields $L_Y$ are also globally-defined on the cylinder and therefore must be linear combinations of $\mathfrak{so}(3,2)$ generators. In this case, conformal symmetry of the CFT$_3$ implies the Ward identity
\begin{eqnarray}
\label{eq:action-corr}
    \sum_{i=1}^n \delta_{L_{Y_i}}\langle{\cal O}_1\cdots{\cal O}_n\rangle = 0.
\end{eqnarray}
In the large $R$ limit this reduces to
\begin{equation}
\begin{split}\label{eq:global-cft2-symmetry}
    \sum_{i=1}^n \left[D_{z_i}Y^{z_i} h_i +D_{\bar z_i}Y^{\bar z_i}\bar{h}_i+Y^{z_i}(\partial_{z_i}- s_i\Omega_{z_i})+Y^{\bar z_i}(\partial_{\bar z_i} -s_i\Omega_{\bar z_i})+O(R^{-1})\right]\langle{\cal O}_1\cdots{\cal O}_n\rangle =0,
    \end{split}
\end{equation}
which corresponds to the global ${\rm SL}(2,\mathbb{C})/\mathbb{Z}_2$ symmetry of the CCFT$_2$ as expected. When $Y$ are not globally defined, we expect the symmetry action on the correlator \eqref{eq:action-corr} to reduce in the large $R$ limit to an insertion of the CCFT$_2$ stress tensor. In the next section we will show that the subleading conformally soft graviton theorem in CCFT and the associated stress tensor Ward identity follow from the flat limit of the CFT$_3$ shadow stress tensor Ward identities. Remarkably, the large-$R$ expansion of the shadow stress tensor Ward identity in CFT$_3$ allows us to also directly recover the \textit{leading} conformally soft graviton theorem.

\section{CCFT$_{d-1}$ conformally soft theorems from CFT$_d$}
\label{sec:conformally-soft}

In this section we describe how soft symmetries in CCFT$_{d-1}$ emerge from the higher-dimensional CFT$_d$ upon dimensional reduction. As a first step, we identify the operators in CFT$_d$ that become conformally soft operators. In particular, we show that the leading conformally soft gluon in CCFT$_{d-1}$ arises in the flat limit\footnote{Defined here as the localization of the operator at $u = 0$ in a time strip $\tau = \tau_0 +  \frac{u}{R}$ of infinitesimal width. As we show in appendix \ref{app:shadow-dim-red} one can equivalently start from the time-Mellin transformed shadow current \eqref{eq:time-Mellin} in the strip and take $\Delta_0 = 1$. In this paper, the flat space limit, while motivated by holography, doesn't require the CFT$_d$ to have a holographic dual.} from a shadow-transformed conserved  current in CFT$_d$. Similarly, the leading and subleading conformally soft gravitons are obtained from the CFT$_d$ stress tensor. 

The relation between soft theorems in $\mathbb{R}^{1, d+1}$ and shadow stress tensor Ward identities in CFT$_d$ was first observed in \cite{Kapec:2017gsg}. Here we combine this general correspondence with the flat space limit to derive CCFT$_{d-1}$ conserved operators (associated instead with soft theorems in $\mathbb{R}^{1,d}$) from CFT$_d$ ones.

Particularly relevant will be the shadow transform of a spin $J$ tensor field in CFT$_d$ which is defined in the embedding space (see appendix \ref{sec:embedding-space}) as
\begin{eqnarray}
\label{eq:embedding-shadow}
\widetilde{\Phi}^{A_1\cdots A_J}(P)\equiv \int D^d Y\dfrac{\prod_{i}(\eta^{A_iB_i}(P\cdot Y)-Y^{A_i}P^{B_i})}{(-2P\cdot Y)^{d-\Delta+J}}{\Phi}_{B_1\cdots B_J}(Y).
\end{eqnarray}
The shadow transform squares to the identity up to normalization \cite{Simmons-Duffin:2012juh}.
This integral transform maps a primary of dimension and spin $(\Delta,J)$ to another primary of dimension and spin $(d-\Delta,J)$. In the remainder of this section we lift the analysis of \cite{Kapec:2017gsg} to the embedding space $\mathbb{R}^{1,d+1}$ and evaluate shadow current and shadow stress tensor insertions
\begin{eqnarray}
    \langle \widetilde{J}_A(P){\cal O}_1(P_1)\cdots{\cal O}_n(P_n)\rangle,\quad \langle \widetilde{T}_{AB}(P){\cal O}_1(P_1)\cdots{\cal O}_n(P_n)\rangle.
\end{eqnarray} 
Our approach is therefore independent on the choice of lightcone section or conformally flat manifold $(\Sigma,g)$.  In order to take the flat space limit we project and analytically continue to CFT$_d$ on the Lorentzian cylinder.
To simplify formulas we introduce the notation $\mathbb{X}$ for a string of primary field insertions in correlation functions
\begin{eqnarray}
  \langle  \mathbb{X} \rangle &\equiv&   \langle {\cal O}_1(P_1)\cdots{\cal O}_n(P_n) \rangle.
\end{eqnarray}

Since the dimensions of the leading conformally soft gluon and subleading conformally soft gravitons are $\Delta = 1$ and $\Delta = 0$ respectively in any number of dimensions, it is perhaps to be expected that the flat limit will lead the corresponding conformally soft theorems. What we find remarkable is that this approach also allows us to easily recover the the leading conformally soft graviton! This can be obtained by acting on the CFT$_{d}$ shadow stress tensor with $\p_u$ in the strip. We will see that in the limit $R \rightarrow \infty$ this indeed precisely reproduces the leading conformally soft graviton theorem in CCFT$_{d-1}$.

\subsection{Shadow current}
\label{sec:embedding-gluon}

Using the defining relation \eqref{eq:embedding-shadow}, the shadow transform of a spin-1 field in the embedding space can be written as
\begin{eqnarray}
\label{eq:current-shadow}
    \widetilde{J}_A(P) &=& \dfrac{1}{4}\int D^dY \dfrac{\partial_{P^A}\partial_{Y^B}\log(-2P\cdot Y)}{(-2P\cdot Y)^{d-\Delta-1}} J^B(Y).
\end{eqnarray}
Here we have used the following identities
\begin{eqnarray}\label{eq:inversion-tensor-embedding}
    \dfrac{\partial}{\partial P^A}\log(-2P\cdot Y) &=& \dfrac{Y_A}{P\cdot Y},\quad
    \dfrac{\partial}{\partial P^A}\dfrac{\partial}{\partial Y^B}\log(-2P\cdot Y) = \dfrac{\eta_{AB}(P\cdot Y)-P_BY_A}{(P\cdot Y)^2}.
\end{eqnarray}

We now consider a $\mathfrak{g}$-valued current where $\mathfrak{g}$ is the Lie algebra of a Lie group $G$ which is a global symmetry of the CFT$_d$. Omitting color indices and recalling that the dimension of a current is $\Delta=d-1$, \eqref{eq:current-shadow} reduces to
\begin{eqnarray}
    \widetilde{J}_A(P) &=& \dfrac{1}{4}\int D^dY \partial_{P^A}\partial_{Y^B}\log(-2P\cdot Y) J^B(Y)\\
   &=& -\dfrac{1}{4}\int D^dY \partial_{P^A}\log(-2P\cdot Y) \partial_{Y^B}J^B(Y),
\end{eqnarray}
where in the last line we have integrated by parts.\footnote{ Recall that on the lightcone $J^B(Y) \sim J^B(Y) + Y^B f(Y)$. }
We now invoke the Ward identity\footnote{The embedding space delta function $\delta(Y,P_i)$ is defined by $\int D^d Y \delta(Y, P_i) = 1$.} \cite{Costa:2011mg}
\begin{eqnarray}
   \partial_B \langle J^B(Y)\mathbb{X}\rangle &=& \sum_{i=1}^n \delta(Y,P_i)T_i\langle\mathbb{X}\rangle,
\end{eqnarray}
where $T_i$ are the generators of the representation of $G$ in which ${\cal O}_i$ transforms. It follows immediately that 
\begin{eqnarray}
\label{eq:current-Ward}
    \langle \widetilde{J}_A(P)\mathbb{X}\rangle &=& -\dfrac{1}{4}\sum_{i=1}^n  \dfrac{(P_i)_A}{P\cdot P_i}T_i\langle\mathbb{X}\rangle.
\end{eqnarray}

Finally, we can project \eqref{eq:current-Ward} to a particular section of the lightcone parameterized by $P^A(x)$. In this case we find
\begin{eqnarray}
    \langle \widetilde{J}_\mu(x){\cal O}_1(x_1)\cdots{\cal O}_n(x_n)\rangle &=& -\dfrac{1}{4}\sum_{i=1}^n \dfrac{\partial_\mu P(x)\cdot P(x_i)}{P(x)\cdot P(x_i)}T_i\langle{\cal O}_1(x_1)\cdots{\cal O}_n(x_n)\rangle.
\end{eqnarray}
Equivalently, as described in appendix \ref{sec:embedding-space} we can choose a set of orthogonal polarization tensors $\varepsilon_a^A(x)$ \eqref{eq:pol-vectors} and project the components of the shadow current to an orthogonal basis obtaining
\begin{eqnarray}\label{eq:shadow-current-final}
    \langle \widetilde{J}_a(x){\cal O}_1(x_1)\cdots{\cal O}_n(x_n)\rangle &=& -\dfrac{1}{4}\sum_{i=1}^n \dfrac{\varepsilon_a(x)\cdot P(x_i)}{P(x)\cdot P(x_i)}T_i\langle{\cal O}_1(x_1)\cdots{\cal O}_n(x_n)\rangle,
\end{eqnarray}
which coincides with the leading soft gluon theorem in the embedding space $\mathbb{R}^{1,d+1}$ with the soft gluon operator given by \cite{Kapec:2017gsg} \footnote{Note that we normalize the shadow transform \eqref{eq:embedding-shadow} according to \cite{Simmons-Duffin:2012juh}. This normalization differs from the one in \cite{Kapec:2017gsg} by a factor of $(-1/2)^J$. To see this, note that when contracted onto lightcone tensors,
\be 
\label{eq:inversion-identity}
\begin{split}
\frac{1}{4}\frac{\eta_{AB}(P\cdot Y) - P_B Y_A}{(P\cdot Y)} J^B(Y) &= \frac{1}{4}\frac{\eta_{AB}(P\cdot Y) - P_B Y_A - Y_B P_A}{(P\cdot Y)} J^B(Y) \\
&= -\frac{1}{2(P - Y)^2}\left[\eta_{AB} - 2\frac{(P - Y)_A (P - Y)_B}{(P - Y)^2} \right]J^B(Y).
\end{split}
\ee
}
\be 
\mathscr{S}_a(x) \equiv  -4 \widetilde{J}_a(x).
\ee 
 Our main result will be to demonstrate that analytic continuation to Lorentzian signature followed by the flat limit prescription of \cite{PipolodeGioia:2022exe} will yield the leading conformally soft gluon. The leading and subleading conformally soft gravitons in CCFT$_{d-1}$ (or equivalently the soft graviton in $\mathbb{R}^{1,d}$) can be recovered in a similar way from the CFT$_d$ stress tensor. To show this, we first need to generalize the embedding space analysis herein to the shadow stress tensor.

\subsection{Shadow stress tensor}
\label{sec:embedding-graviton}

For a spin two field the shadow transform takes the form
\begin{eqnarray}
\label{eq:shadow-stress}
    \widetilde{T}_{AB}(P) &=& \dfrac{1}{16}\int D^dY \dfrac{\partial_{P^A}\partial_{Y^C}\log(-2P\cdot Y)\partial_{P^B}\partial_{Y^D}\log(-2P\cdot Y)}{(-2P\cdot Y)^{d-\Delta-2}}T^{CD}(Y).
\end{eqnarray}
For the stress tensor, $\Delta=d$ and so
\begin{equation}
    \widetilde{T}_{AB}(P) = \dfrac{1}{16}\int D^dY (-2P\cdot Y)^2\partial_{P^A}\partial_{Y^C}\log(-2P\cdot Y)\partial_{P^B}\partial_{Y^D}\log(-2P\cdot Y)T^{CD}(Y).
\end{equation}
While the steps involved in the derivation of the relation between the shadow transform of the stress tensor and the soft graviton theorem are similar to those in \cite{Kapec:2017gsg}, we find it instructive to repeat the significantly simpler calculation here in the embedding space. Integrating by parts and using \eqref{eq:inversion-tensor-embedding} this can be written as
\begin{eqnarray}
    \widetilde{T}_{AB}(P) 
    &=&-\dfrac{1}{8}\int D^dY \dfrac{Y_A}{P\cdot Y}\partial_{Y^C}\left\{[\eta_{BD}(P\cdot Y)-P_D Y_B]T^{CD}(Y)\right\} +  (A \leftrightarrow B)
\end{eqnarray}
and further evaluating the derivative with respect to $Y$ one finds
\begin{eqnarray}\label{eq:shadow-stress-tensor-1}
    \widetilde{T}_{AB}(P) &=& \dfrac{1}{4}\int D^dY \dfrac{Y_A}{P\cdot Y}\eta_{B[C}P_{D]}T^{CD}(Y)\nonumber\\
    &-& \dfrac{1}{8}\int D^dY \dfrac{Y_A}{P\cdot Y} [\eta_{BD}(P\cdot Y)-P_D Y_B]\partial_{Y^C}T^{CD}(Y) + (A \leftrightarrow B),
\end{eqnarray}
where $[. ,.]$ stands for antisymmetrization.  We ensured that the manifest symmetry of \eqref{eq:shadow-stress} under $A \leftrightarrow B$ is preserved upon integration by parts.

The insertions of both terms on the RHS of \eqref{eq:shadow-stress-tensor-1} in correlation functions are determined by the uplift of the stress tensor Ward identities  to the embedding space \cite{Costa:2011mg}. In particular, the first line involves $T^{[CD]}$ whose insertions are related to the spin component  ${\cal S}^{CD}$ of the Lorentz generators in the embedding space 
\begin{eqnarray}
    \langle T^{[CD]}(Y)\mathbb{X}\rangle &=& -\dfrac{i}{2}\sum_{i=1}^n \delta(Y,P_i){\cal S}_i^{CD}\langle\mathbb{X}\rangle.
\end{eqnarray}
We then find that inside correlation functions, the first line in \eqref{eq:shadow-stress-tensor-1} simplifies to
\begin{eqnarray}
    \dfrac{1}{4}\int D^dY \dfrac{Y_A}{P\cdot Y}\eta_{B[C}P_{D]}\langle T^{CD}(Y)\mathbb{X}\rangle &=&-\dfrac{i}{8}\sum_{i=1}^n \dfrac{(P_i)_A P_D}{P\cdot P_i}\eta_{BC}{\cal S}_i^{CD}\langle\mathbb{X}\rangle\nonumber\\
    &=&\dfrac{i}{8}\sum_{i=1}^n \dfrac{(P_i)_A P^D}{P\cdot P_i}({\cal S}_i)_{DB}\langle\mathbb{X}\rangle.
\end{eqnarray}
On the other hand, the second term in \eqref{eq:shadow-stress-tensor-1} is determined by the stress tensor Ward identity
\begin{eqnarray}
    \langle \partial_{Y^C}T^{CD}(Y)\mathbb{X}\rangle &=& -\eta^{DE}\sum_{i=1}^n \delta(Y,P_i)\partial_{P_i^E}\langle \mathbb{X}\rangle.
\end{eqnarray}
Using this Ward identity, insertions of the second term in \eqref{eq:shadow-stress-tensor-1} can then be shown to be related to the orbital part of the embedding space Lorentz generators, ${\cal L}_{DB}$, namely 
\begin{eqnarray}
    {\cal L}_{DB} &\equiv& -i(P_D \partial_{P^B}-P_B\partial_{P^D}).
\end{eqnarray}

Specifically, we find that inside correlation functions the second term in \eqref{eq:shadow-stress-tensor-1} reduces to
\begin{equation}
    -\dfrac{1}{8}\int D^dY \dfrac{Y_A}{P\cdot Y}[\eta_{BD}(P\cdot Y)-P_D Y_B]\langle\partial_{Y^C}T^{CD}(Y)\mathbb{X}\rangle = \dfrac{i}{8}\sum_{i=1}^n\dfrac{(P_i)_A P^D}{P\cdot P_i}({\cal L}_i)_{DB}\langle \mathbb{X}\rangle.
\end{equation}
Combining the two contributions from equation \eqref{eq:shadow-stress-tensor-1} we find the embedding space formula for insertions of the stress tensor in CFT$_d$
\begin{eqnarray}
\label{eq:embedding-Ward}
    \langle \widetilde{T}_{AB}(P)\mathbb{X}\rangle &=& \dfrac{i}{8}\sum_{i=1}^n \dfrac{(P_i)_A P^D}{P\cdot P_i}[({\cal L}_i)_{DB}+({\cal S}_i)_{DB}]\langle \mathbb{X}\rangle + (A \leftrightarrow B)\nonumber\\
    &\equiv&\dfrac{i}{8}\sum_{i=1}^n \dfrac{(P_i)_A P^D}{P\cdot P_i}({\cal J}_i)_{DB}\langle \mathbb{X}\rangle +  (A \leftrightarrow B).
\end{eqnarray}
As before, we can now project to a particular section parameterized by  $P^A(x)$
\begin{eqnarray}
    \langle \widetilde{T}_{\mu\nu}(x)\mathbb{X}\rangle &=& \dfrac{\partial P^A}{\partial x^{\mu}}\dfrac{\partial P^B}{\partial x^{\nu}}\langle \widetilde{T}_{AB}(P(x))\mathbb{X}\rangle \nonumber\\
    &=&\dfrac{i}{4}\sum_{i=1}^n \dfrac{\partial_{{\{}\mu} P^A(x)\partial_{\nu{\}}} P^B(x) P_A(x_i) P^D(x)}{P(x)\cdot P(x_i)}({\cal J}_i)_{DB}\langle \mathbb{X}\rangle.
\end{eqnarray}

Alternatively, using the orthogonal set of polarization vectors $\varepsilon_a^A$  \eqref{eq:pol-vectors} to construct the spin two tensors $\varepsilon_{ab}^{AB} = \varepsilon_{\{a}^A\varepsilon_{b\}}^B$ and projecting to the associated orthonormal basis, we find \cite{Kapec:2017gsg}
\begin{equation}\label{eq:shadow-stress-tensor-final}
    \langle \widetilde{T}_{ab}(x){\cal O}_1(x_1)\cdots{\cal O}_n(x_n)\rangle =\dfrac{i}{4}\sum_{i=1}^n \dfrac{\varepsilon_{ab}^{AB}(x) P_A(x_i) P^D(x)}{P(x)\cdot P(x_i)}({\cal J}_i)_{DB}\langle {\cal O}_1(x_1)\cdots{\cal O}_n(x_n)\rangle,
\end{equation}
which upon defining\footnote{Working in units where $\kappa = \sqrt{32\pi G} = 2.$}
\be 
\mathscr{G}_{ab} = -4\widetilde{T}_{ab}
\ee 
we recognize as the formula for a subleading soft graviton insertion in the embedding space $\mathbb{R}^{1,d+1}$.

\subsection{Large $R$ expansions}
\label{sec:flat-space}

We now apply these results to a CFT$_d$ on the Lorentzian cylinder and show that the conformally soft theorems in the dimensionally reduced CCFT$_{d-1}$ arise naturally from the flat space limit prescription proposed in \cite{PipolodeGioia:2022exe}. We work with the analytic continuation to Lorentzian signature of the Euclidean results derived in the previous sections.

Consider the embedding 
\begin{eqnarray}
\label{cylinder-embedding}
    P(\tau,\vec{z}) &=& (\sin\tau,\Omega(\vec{z}),\cos\tau)
\end{eqnarray}
of the $d$-dimensional Lorentzian cylinder in $\mathbb{R}^{2,d}$ with metric $\eta_{AB} = (-1, 1, \cdots, -1)$ introduced in section \ref{sec:preliminaries}. Here $\Omega^2 = 1$ are unit normals to $S^{d - 1}$.
We also consider the polarization tensors
\begin{eqnarray}
\label{eq:pol-cylinder}
    \varepsilon_{a}(\tau,\vec{z}) &=& (z_a\sin\tau,\delta_a^b,-z_a,z_a\cos\tau),\quad a=1,\dots,d-1,\\
    \varepsilon_{d}(\tau,\vec{z}) &=& (\cos\tau,\vec{0},-\sin\tau),
\end{eqnarray}
where $\delta^b_a$ denotes a vector with vanishing components except for an entry equal to 1 at $b = a$.
These are such that $\varepsilon_{a}\cdot P = \varepsilon_d\cdot P=0$ provided that
\be 
z_a = \frac{\Omega_a}{1 + \Omega_d}, \quad a = 1, \cdots d-1.
\ee
Moreover, $\varepsilon_a\cdot \varepsilon_b = \eta_{ab}$ where $\eta_{dd}=-1$. They also enjoy the property that setting $\tau =  \frac{\pi}{2}+\frac{u}{R}$ and expanding at large $R$
\begin{equation}
\label{eq:flat-limit-pol}
\begin{split}
    \varepsilon_a &= (z_a,\delta_a^b,-z_a, 0)+O(R^{-1}),\\
    \varepsilon_d &= (0,\vec{0},-1)+O(R^{-1}).
    \end{split}
\end{equation}
We therefore see that $\varepsilon_a = (\epsilon_a,0)+O(R^{-1})$ where $\epsilon_a$ are polarization  vectors in $\mathbb{R}^{1,d}$ \cite{Kapec:2017gsg}.
In the case of CFT$_3$ ($d = 3$), it will be convenient to trade the coordinates $(z^1,z^2)$ for complex coordinates $(z,\bar z) \equiv (z_1 + iz_2, z_1 - iz_2)$, and $\varepsilon_1(\tau,\vec{z})$ and $\varepsilon_2(\tau,\vec{z})$ for the following linear combinations
\begin{eqnarray}
\label{eq:helicities}
    \varepsilon_z(\tau,z,\bar z) = \frac{1}{{\sqrt{2}}}(\bar z\sin\tau,1,-i,-\bar z,\bar z\cos\tau),\quad \varepsilon_{\bz}(\tau,z,\bar z) = \frac{1}{{\sqrt{2}}}(z\sin\tau,1,i,-z,z\cos\tau).
\end{eqnarray}
In the flat space limit, \eqref{eq:helicities} become $\varepsilon_{a} = (\epsilon_a,0)+O(R^{-1})$ with $\epsilon_z$ and $\epsilon_{\bz}$ the polarization vectors associated respectively with positive and negative helicities in $\mathbb{R}^{1,3}$, namely
\begin{eqnarray}
    \epsilon_z(z,\bz) &=& \dfrac{1}{\sqrt{2}}(\bz,1,-i,-\bz),\quad \epsilon_{\bz}(z,\bz) = \dfrac{1}{\sqrt{2}}(z,1,i,-z).
\end{eqnarray}

For simplicity we will assume that all of the operators are placed at $\tau = \frac{\pi}{2}$, which holographically would amount to considering all bulk particles to be outgoing. If one of the particles is taken to be incoming, following \cite{PipolodeGioia:2022exe} we insert the corresponding operator at $(-\frac{\pi}{2},\vec{z}^A)$ where $\vec{z}^A$ denotes the antipodal map. In that case we observe that $P(-\frac{\pi}{2},\vec{z}^A) = - P(\frac{\pi}{2},\vec{z})$. Taking this into account therefore produces the required sign difference in the corresponding contribution to the leading soft graviton factor.
Finally, recall that at large $R$ and $\tau = \frac{\pi}{2} + \frac{u}{R}$
\begin{eqnarray}
\label{eq:flat-limit-mom}
    P(\tau,\vec{z}) &=& (q(\vec{z}),0)+O(R^{-1}),
\end{eqnarray}
where $q(\vec{z}) = (1,\Omega(\vec{z}))$ is a null vector in $\mathbb{R}^{1,d}$.

\subsubsection{Leading conformally soft gluon theorem}
\label{sec:gluons}

Equipped with these results,  consider a ${\frak g}$-valued conserved current $J$ in a CFT$_d$ with global symmetry group $G$. Insertions of the shadow transform of this current into correlation functions on the Lorentzian cylinder are obtained from the embedding space formula \eqref{eq:shadow-current-final} by projecting with the polarization tensors $\{\varepsilon_a,\varepsilon_d\}$  in \eqref{eq:pol-cylinder}.
Expanding at large $R$ and using \eqref{eq:flat-limit-pol} together with \eqref{eq:flat-limit-mom} we find
\begin{eqnarray}
    \langle \mathscr{S}_a(x){\cal O}_1(x_1)\cdots{\cal O}_n(x_n)\rangle &=& \sum_{i=1}^n \dfrac{\epsilon_a(x)\cdot q(x_i)}{q(x)\cdot q(x_i)}T_i\langle{\cal O}_1(x_1)\cdots{\cal O}_n(x_n)\rangle+O(R^{-1}),
\end{eqnarray}
which reproduces the leading conformally soft gluon theorem in CCFT$_{d-1}$. Note that in the limit $u \rightarrow 0$ the large $R$ corrections drop out. In the particular case of CFT$_3$ using the set of polarizations $\{\varepsilon_{z},\varepsilon_{\bz},\varepsilon_3\}$ we find
\begin{eqnarray}\label{eq:current-soft-factors-cft3-ccft2}
    \dfrac{\epsilon_z(x)\cdot q(x_i)}{q(x)\cdot q(x_i)} &=& \frac{1}{\sqrt{2}}\dfrac{1+z\bar z}{z-z_i},\quad \dfrac{\epsilon_{\bz}(x)\cdot q(x_i)}{q(x)\cdot q(x_i)} = \frac{1}{\sqrt{2}} \dfrac{1+z\bar z}{\bar z-\bar z_i},
\end{eqnarray}
and therefore we recover
\begin{eqnarray}
\label{eq:soft-gluon}
    \langle \mathscr{S}_z(x){\cal O}_1(x_1)\cdots{\cal O}_n(x_n)\rangle &=& \frac{1+z\bar z}{\sqrt{2}}\sum_{i=1}^n \dfrac{T_i}{z-z_i}\langle{\cal O}_1(x_1)\cdots{\cal O}_n(x_n)\rangle+O(R^{-1}),\\
    \langle \mathscr{S}_{\bz}(x){\cal O}_1(x_1)\cdots{\cal O}_n(x_n)\rangle &=& \frac{1+z\bar z}{\sqrt{2}}\sum_{i=1}^n \dfrac{T_i}{\bz-\bz_i}\langle{\cal O}_1(x_1)\cdots{\cal O}_n(x_n)\rangle+O(R^{-1}),\label{eq:soft-gluon1}
\end{eqnarray}
which are the holomorphic and antiholomorphic $\mathfrak{g}$-Kac-Moody Ward identities \cite{Strominger:2013lka}. 

The time component of the CFT$_3$ shadow current leads to an identity that resembles a soft scalar theorem \cite{DiVecchia:2015jaq}
\be 
\label{eq:u-comp}
\langle \widetilde{J}_u(x) \mathcal{O}_1(x_1) \cdots \mathcal{O}_n(x_n) \rangle \sim \frac{u}{R}\sum_{i = 1}^n \frac{T_i}{q(x) \cdot q(x_i)}  \langle \mathcal{O}_1(x_1) \cdots \mathcal{O}_n(x_n) \rangle + O(R^{-3}).
\ee
Note that the leading term in \eqref{eq:u-comp} is of a different order in a large $R$ expansion compared to \eqref{eq:soft-gluon}, \eqref{eq:soft-gluon1}. Such soft theorems were argued in \cite{Campiglia:2017dpg, Campiglia:2018see} to arise from conservation laws associated with higher form symmetries in 4D AFS. From a boundary perspective, we find that they are a simple consequence of dimensional reduction. It would be interesting yet beyond the scope of this paper to understand the relation between these different perspectives, as well as the role of these additional symmetries in CCFT$_{d-1}$.

\subsubsection{Leading and subleading conformally soft graviton theorems}
\label{sec:gravitons}

Next we consider the shadow stress tensor $\widetilde{T}_{AB}(P)$ whose insertions are given by \eqref{eq:embedding-Ward} or, upon projection to the Lorentzian cylinder, by \eqref{eq:shadow-stress-tensor-final}. As we show in details in Appendix \ref{app:soft-graviton-theorems} restricting to components on a constant time slice $a,b\in \{1,\dots, d-1\}$, we find in the flat limit that 
\begin{eqnarray}
\label{eq:leading-soft}
	\p_u \langle \mathscr{G}_{\{ab\}}{\cal O}_1\cdots{\cal O}_n\rangle &=& \sum_{i=1}^n \dfrac{\epsilon_{ab}^{AB}(x)q_A(x_i)q_B(x_i)}{q(x)\cdot q(x_i)}\p_{u_i}\langle{\cal O}_1\cdots{\cal O}_n\rangle + O(R^{-1}).
\end{eqnarray}
Here $\epsilon_{ab}$ is the transverse, traceless polarization tensor in $\mathbb{R}^{1,d}$.
Upon switching to a basis that diagonalizes the dilatation operator on $S^{d-1}$ via the transform \eqref{eq:time-Mellin}, $\partial_{u_i}$ becomes the weight-shifting operator $e^{\partial_{\Delta_i}}$. Note that in the limit $u \rightarrow 0$, the large $R$ corrections to \eqref{eq:leading-soft} drop out.  We hence see that insertions of $\underset{u \rightarrow 0}{\rm lim}\p_u\mathscr{G}_{{\{ab\}}}$ reproduce the leading conformally soft graviton theorem in $\mathbb{R}^{1,d}$ with $\mathscr{N}^{(0)}_{ab} \equiv {\underset{u \rightarrow 0}{\rm lim}}\p_u \mathscr{G}_{\{ab\}}$ the leading soft graviton operator. 

Moreover, we show in Appendix \ref{app:soft-graviton-theorems}, that 
\begin{equation}
\label{eq:subleading-soft}
	(1-u\p_u)\langle \mathscr{G}_{\{ab\}}{\cal O}_1\cdots{\cal O}_n\rangle = i\sum_{i=1}^n \dfrac{\epsilon_{ab}^{AB}(x)q_A(x_i)q^C(x_i)}{q(x)\cdot q(x_i)}({\cal J}_i)_{BC}\langle{\cal O}_1\cdots{\cal O}_n\rangle+O(R^{-1}),
\end{equation}
where $({\cal J}_i)_{BC}$ have indices restricted to $B,C<d+1$ due to $\epsilon_a^{d+1}=q^{d+1}=0$. In this case, $({\cal J}_i)_{BC}$ coincide with the $\mathfrak{so}(d,2)$ generators whose action on conformal primary operators restricted to the strip \eqref{metric-exp} was worked out in section \ref{sec:ckv-action}. Their action hence coincides with that of the Lorentz generators in $(d+1)$-dimensional AFS, or equivalently, conformal $\mathfrak{so}(d,1)$ transformations. Therefore insertions of ${\underset{u \rightarrow 0}{\rm lim}}(1-u\p_u)\mathscr{G}_{\{ab\}}$ reproduce the subleading conformally soft graviton theorem in $\mathbb{R}^{1,d}$  and the subleading conformally soft graviton operator is related to the CFT$_d$ shadow stress tensor via $\mathscr{N}^{(1)}_{ab} \equiv {\underset{u \rightarrow 0}{\rm lim}}(1-u\p_u)\mathscr{G}_{\{ab\}}$. The constructions of the supertranslation current and the stress tensor from $\mathscr{N}_{ab}^{(0)}$ and $\mathscr{N}_{ab}^{(1)}$ then follow directly from  respectively \cite{as1,He:2014laa} and \cite{Kapec:2014opa,Kapec:2016jld} .

We now specialize to CFT$_3$. Using the large $R$ expansions \ref{eq:flat-limit-pol} of the polarization tensors $\{\varepsilon_z,\varepsilon_{\bz},\varepsilon_3\}$ we construct the transverse traceless spin 2 polarization tensors $\epsilon_{ab} = \epsilon_{\{a}\epsilon_{b\}}$. The only non-vanishing components are $\epsilon_{zz}^{AB} =\epsilon_z^A\epsilon_z^B$ and $\epsilon_{\bz\bz}^{AB}=\epsilon_{\bz}^A\epsilon_{\bz}^B$. Therefore the expressions for the leading soft factors reduce to those derived in \cite{He:2014laa},
\begin{eqnarray}
	\dfrac{\epsilon_{zz}^{AB}(x)q_A(x_i)q_B(x_i)}{q(x)\cdot q(x_i)} &=&  -\dfrac{\bz-\bz_i}{z-z_i}\dfrac{1+z\bz}{1+z_i\bz_i},\\
	\dfrac{\epsilon_{\bz\bz}^{AB}(x)q_A(x_i)q_B(x_i)}{q(x)\cdot q(x_i)} &=& -\dfrac{z-z_i}{\bz-\bz_i}\dfrac{1+z\bz}{1+z_i\bz_i},
\end{eqnarray}
and consequently
\begin{eqnarray}
	\langle \mathscr{N}_{zz}^{(0)}{\cal O}_1\cdots{\cal O}_n\rangle &=& -\sum_{i=1}^n \dfrac{\bz-\bz_i}{z-z_i}\dfrac{1+z\bz}{1+z_i\bz _i}\p_{u_i}\langle{\cal O}_1\cdots{\cal O}_n\rangle ,\\
	\langle \mathscr{N}_{\bz\bz}^{(0)}{\cal O}_1\cdots{\cal O}_n\rangle &=& -\sum_{i=1}^n \dfrac{z-z_i}{\bz-\bz_i}\dfrac{1+z\bz}{1+z_i\bz _i}\p_{u_i}\langle{\cal O}_1\cdots{\cal O}_n\rangle .
\end{eqnarray}

Insertions of $\mathscr{N}^{(1)}_{zz}$ and $\mathscr{N}^{(1)}_{\bz\bz}$ can be treated similarly. Relegating the complete calculation to Appendix \ref{app:subleading-soft-factor-ccft2}, we find that
\begin{equation}
\begin{split}
	\langle \mathscr{N}_{zz}^{(1)}{\cal O}_1\cdots{\cal O}_n\rangle &= \sum_{i=1}^n \left[\dfrac{(\bz-\bz_i)(1+\bz z_i)}{( z- z_i)(1+z_i\bar z_i)}2\bar{\mathfrak{h}}_i-\dfrac{(\bz- \bz_i)^2}{ z- z_i}(\partial_{\bz_i} -\Omega_{\bz_i}J_3)\right]\langle{\cal O}_1\cdots{\cal O}_n\rangle,\\
\langle \mathscr{N}_{\bz\bz}^{(1)}{\cal O}_1\cdots{\cal O}_n\rangle &= \sum_{i=1}^n \left[\dfrac{(z-z_i)(1+z \bar z_i)}{(\bar z-\bar z_i)(1+z_i\bar z_i)}2{\mathfrak{h}}_i-\dfrac{(z- z_i)^2}{\bar z-\bar z_i}(\partial_{z_i} -\Omega_{z_i}J_3)\right]\langle{\cal O}_1\cdots{\cal O}_n\rangle,
\end{split}
\end{equation}
 which agrees with the formula for the subleading soft factor \cite{Kapec:2014opa, Kapec:2016jld} with external weights $(\mathfrak{h}_i,\mathfrak{h}_i)$ and helicities $J_3$ as defined in \eqref{eq:weights}. Taking a two-dimensional shadow transform of $\mathscr{N}_{ab}^{(1)}$ as in \cite{Kapec:2016jld} yields the CCFT$_2$ stress tensor. 

\section{Discussion}

In this paper we studied the symmetries of CFT$_3$ on the Lorentzian cylinder over short time intervals. We showed that strips of infinitesimal width $\propto R^{-1}$ around any time-slice admit an infinite-dimensional set of locally-defined solutions in the $R\to \infty$ limit. These can be reorganized into vector fields obeying the $\mathfrak{ebms}_4$ algebra. The extended BMS$_4$ symmetry emerges via a Inonu-Wigner contraction which for the global subalgebra reduces to the contraction of the  $\mathfrak{so}(3,2)$ algebra to Poincar\'e. We  studied the transformation properties of CFT$_3$ primary operators in the strip under the superrotation subalgebra of $\mathfrak{ebms}_4$  and found that they transform as two-dimensional conformal primaries with operator-valued effective dimensions $\hat{\Delta} = \Delta+u\p_u$.

The two-dimensional dilatation can be diagonalized by a time Mellin-like transform.  Consequently each CFT$_3$ primary operator results in a continuum of CCFT$_2$ primary operators of the same spin and with dimensions $\Delta_{\rm CCFT} = \Delta + \Delta_0-1$ where $\Delta$ is the CFT$_3$ dimension and $\Delta_0$ is the dual Mellin dimension. We argued that the special case $\Delta_0 = 1$ implements a restriction to the $u=0$ time-slice, in agreement with previous results \cite{PipolodeGioia:2022exe}.

We showed that, inside the strip, the transverse components $\widetilde{T}_{ab}$ of the $\Delta=0$ shadow stress tensor give rise to operators $\mathscr{N}_{ab}^{(0)}$ and $\mathscr{N}_{ab}^{(1)}$ whose insertions into correlation functions reproduce the leading and subleading conformally soft graviton theorems. Likewise, the transverse components $\widetilde{J}_a$ of the $\Delta =1$ shadow current provide an operator $\mathscr{S}_a$ whose insertions reproduce the leading soft gluon theorem. As such, conformally soft theorems and the corresponding infinite-dimensional CCFT$_{d-1}$ symmetries effectively emerge from the dimensional reduction of the CFT$_d$.

There are several aspects of our dimensional reduction or flat space limit that we believe deserve further investigation. The conformal Killing vectors \eqref{asy-ckv} giving rise to the $\mathfrak{ebms}_4$ algebra violate the conformal Killing equation at finite $R$. This appears to be in stark contrast to the asymptotic symmetries of 4D AFS that are exact and can be extended into the bulk. It would be interesting to understand  whether the symmetries can be preserved in the strip beyond the $R \rightarrow \infty$ limit and relate this to the emergence of a bulk radial direction from the CFT. Interestingly, both large $r$ corrections to the asymptotic charges in 4D AFS and corrections away from the large AdS radius limit have been linked to loop corrections \cite{Campiglia-soft-th,Banerjee:2022oll}. It would also be interesting to connect our enhanced conformal Killing symmetries \eqref{asy-ckv} in the strip to the bulk $\Lambda$-BMS algebra \cite{Compere:2019bua} which similarly arises, subject to certain boundary conditions, in the limit of infinite AdS radius. 

More generally, our analysis provides motivation for looking for boundary conditions in AdS that turn on shadow operators on the boundary. These operators are dual to modes in AdS that are in general non-normalizable near the boundary, but normalizable deep inside the bulk. This seems consistent with the flat space limit prescription which amounts to zooming in close to the center of AdS \cite{Hijano:2019flat,Hijano:2020szl}, as well as proposals suggesting that flat space physics may be obtained via a $T\bar{T}$ deformation \cite{Giveon:2017nie,Asrat:2017tzd}. It would also be interesting to understand if the whole tower of $w_{1+\infty}$ currents in celestial CFT \cite{Strominger:2021lvk} can similarily arise from a limit of CFT$_3$.

The approach we have adopted in this paper proposes a connection between CCFT and standard CFT. In principle these ideas may allow for an understanding of how general features of CFT, such as the existence of an associative OPE, are reflected in the dimensionally reduced theory, potentially allowing for a better understanding of the corresponding features of CCFT. In particular, our results suggest that the stress tensor of the reduced theory is closely related to the stress tensor of the parent CFT, so that it may be possible to extract a CCFT central charge from this procedure. This may shed light on previous proposals  based on a hyperbolic slicing of Minkowski spacetime \cite{Cheung:2016iub, Pasterski:2022lsl, Ogawa:2023wedge}.

Finally, the shadow transform played an important role in this analysis, since it allowed for the construction of the soft operators from the stress tensor and current. In Lorentzian signature, the shadow transform constructed by Wick rotating the Euclidean shadow is just one member out of a group of transformations preserving the Casimirs of the conformal group \cite{Kravchuk:2018htv}. It therefore seems plausible that the other transforms will also play meaningful roles in the dimensionally reduced CCFT. We hope to address some of these issues in future work.

\section*{Acknowledgements}

We thank Jos\'e Abdalla Helayël-Neto, David Berenstein, Jan de Boer, Jackie Caminiti, Laurent Freidel, Jaume Gomis, Charles Marteau, Walker Melton, Rob Myers, Sruthi Narayanan, Dominik Neuenfeld, Jo\~ao Paulo Pitelli Manoel and Romain Ruzziconi  for discussions. We thank Geoffrey Compère, Johanna Erdmenger, Laurent Freidel, Dan Kapec and Dominik Neuenfeld for comments on a draft. L.P.G. is supported by Conselho Nacional de Desenvolvimento Cient\'{i}fico e Tecnol\'{o}gico (CNPq, process number 140725/2019-9). A.R. is supported by the Heising-Simons Foundation “Observational Signatures of Quantum Gravity” collaboration grant 2021-2817.

\appendix

\section{Embedding space primer}
\label{sec:embedding-space}

A Euclidean CFT$_d$ is defined on the projective null cone in the embedding space $\mathbb{R}^{1,d+1}$ with metric $\eta_{AB}$.\footnote{Lorentzian CFT$_d$ are instead lifted to $\mathbb{R}^{2,d}$.}  The projective null cone is parametrized by a vector $P$ obeying
\begin{eqnarray}
    P^2 &=&0,\quad P\sim \lambda P,\quad \lambda \neq 0.
\end{eqnarray}
 Choosing a representative from each equivalence class yields a section of the lightcone $\Sigma\subset \mathbb{R}^{1,d+1}$ corresponding to a conformally flat manifold on which the CFT$_d$ is realized. The non-linear action of the conformal group on $\Sigma$ is realized through the combination of Lorentz transformations SO$(d+1,1)$ and rescalings of the null cone that preserves the chosen section. Let $P(x)$ be an embedding of $\Sigma$ into $\mathbb{R}^{1,d+1}$. Then the metric it inherits from the ambient space is
\begin{eqnarray}
    ds^2_\Sigma = \eta_{AB}\dfrac{\partial P^A}{\partial x^\mu}\dfrac{\partial P^B}{\partial x^\nu}dx^\mu dx^\nu.
\end{eqnarray}
A different section $\Sigma'$ embedded by $P'(x')$ is related to $\Sigma$ by a rescaling 
\begin{eqnarray}
    P'(x') = \omega(x)P(x).
\end{eqnarray}
The metrics on the two sections $\Sigma, \Sigma'$ can then be shown to be related by a Weyl rescaling
\begin{eqnarray}
    ds^2_{\Sigma'} = \omega^2(x) ds^2_\Sigma.
\end{eqnarray}
We conclude that conformal maps between different conformally flat manifolds are represented in the embedding space by Weyl rescalings  and Lorentz transformations of the embeddings of the corresponding lightcone sections (see \cite{Osborn2019:lectures} for a review). 

A primary field of dimension $\Delta$ and spin $J$ in a CFT$_d$ on a given section can be lifted to a field on the lightcone as follows. If $\phi_{\mu_1\cdots \mu_J}(x)$ is a spin $J$ symmetric traceless tensor, its lift to a tensor $\Phi_{A_1\cdots A_J}(P)$ defined on the embedding space lightcone has to obey the following properties \cite{Costa:2011mg}
\begin{enumerate}
    \item $\Phi_{A_1\cdots A_J}(P)$ is symmetric, traceless and transverse $P^{A_i}\Phi_{A_1\cdots A_J}(P)=0$,

    \item $\Phi_{A_1\cdots A_J}(P)$ is defined up to terms $P_{A_i} \Lambda_{A_1\cdots \hat{A}_i\cdots A_J}(P)$, where $\hat{A}_i$ denotes a missing index,

    \item $\Phi_{A_1\cdots A_J}(P)$ is homogenous of degree $-\Delta$: $\Phi_{A_1\cdots A_J}(\omega P)=\omega^{-\Delta}\Phi_{A_1\cdots A_J}(P)$. 
\end{enumerate}
If $\Sigma$ is parameterized by $P(x)$, $\phi_{\mu_1\cdots \mu_J}(x)$ is then recovered by the projection \cite{Costa:2011mg}
\begin{eqnarray}
    \phi_{\mu_1\cdots \mu_J}(x) = \dfrac{\partial P^{A_1}}{\partial x^{\mu_1}}\cdots \dfrac{\partial P^{A_J}}{\partial x^{\mu_J}}\Phi_{A_1\cdots A_J}(P(x)).
\end{eqnarray}
Projecting using the Jacobian of the embedding as done above reproduces the coordinate components of the tensor field. Alternatively, we can introduce a set of polarization vectors $\varepsilon_a^A(x)$ in the embedding space obeying
\begin{eqnarray}
    \varepsilon_a\cdot P &=&0,\quad \varepsilon_a\cdot \varepsilon_b = \delta_{ab}.
\end{eqnarray}
The pullback of $\varepsilon_a$ to the section $(\Sigma,g)$ can then be shown to give rise to a vielbein in $(\Sigma,g)$, namely
\begin{eqnarray}
\label{eq:pol-vectors}
    e^a_\mu = \dfrac{\partial P^A}{\partial x^\mu}\varepsilon^{a}_A, \quad  \varepsilon^A_a = e^{\mu}_a \frac{\p P^A}{\p x^{\mu}} - (\varepsilon_a\cdot \bar q)q^A, \quad 
\end{eqnarray}
 where \cite{Osborn2019:lectures}
\be\label{eq:completeness-relation}
g^{\mu\nu}\dfrac{\partial P^A}{\partial x^\mu}\dfrac{\partial P^B}{\partial x^\nu} = \eta^{AB}+q^A \bar q^B+ q^B\bar q^A,
\ee
with $g_{\mu\nu} = (P^+)^2\eta_{\mu\nu},$ $q^A = P^A/P^+$ and $\bar q^A = -2\delta^A_-$.

As a result, the symmetric, traceless combination $\varepsilon_{a_1\cdots a_J}^{A_1\cdots A_J} = \varepsilon_{\{a_1}^{A_1}\cdots \varepsilon_{a_J\}}^{A_j}$ can be used as projectors which allow us to recover the components of the tensor field with respect to the orthonormal basis 
\begin{eqnarray}
    \phi_{a_1\cdots a_J}(x) &=&  \varepsilon^{A_1\cdots A_J}_{a_1\cdots a_J}(x)\Phi_{A_1\cdots A_J}(P(x)).
\end{eqnarray}
Primary fields in more general representations of ${\rm SO}(d)$ can be handled in the same way. They are lifted to fields in representations of ${\rm SO}(1,d+1)$ defined on the lightcone with homogeneity of degree $-\Delta$ which are transverse in the appropriate sense and which can be projected back to the original representation by introducing appropriate projection matrices. These fields are again only defined modulo terms that lie in the kernel of the projection matrices. The particular case of Dirac spinors in several dimensions is discussed for example in \cite{Osborn2019:lectures}.

It will also be useful to recall the definition of conformal integrals on the space of homogeneous functions $f(X)$ of degree $-d$  on the lightcone  \cite{Simmons-Duffin:2012juh}
\begin{eqnarray}
    \int D^dXf(X) &=& \dfrac{1}{{\rm Vol}({\rm GL}(1,\mathbb{R})^+)}\int d^{d+2}X \delta(X^2)f(X).
\end{eqnarray}
In practice such integrals are evaluated by gauge-fixing the rescaling freedom and introducing an appropriate Faddeev-Popov determinant.

\section{Properties of the spin covariant derivative}
\label{example}

In this section we show that the spin-covariant derivative \eqref{eq:spin-cov-der} reduces to the Levi-Civita connection when acting on fields transforming in the vector representation of $SO(3)$, namely if
\be 
\left(S_{ab}\right)^{c}_{~d} = -i\left(\delta_{a}^{c}\delta_{bd} - \delta_{ad} \delta_{b}^c \right)
\ee
then
\be 
\nabla_\mu V^\nu =\partial_\mu V^\nu+\Gamma^\nu_{\mu\sigma}V^\sigma.
\ee
To see this we evaluate $\nabla_\mu V^a$ where $V^a$ are the vielbein components of the vector field, and then transform to the coordinate components $\nabla_\mu V^\nu$. We start with 
\begin{eqnarray}
    \nabla_\mu V^a = \partial_\mu V^a+\omega_{\mu\phantom a b}^{\phantom \mu a}V^b.
\end{eqnarray}
The coordinate components are defined by 
\begin{eqnarray}
    \nabla_\mu V^\nu\equiv e^\nu_a \nabla_\mu V^a.
\end{eqnarray}
Evaluating $\nabla_\mu V^\nu$,
\begin{eqnarray}
    \nabla_\mu V^\nu=e_a^\nu \partial_\mu V^a+e_a^\nu\omega_{\mu\phantom a b}^{\phantom \mu a}V^b.
    \end{eqnarray}
We now transform $V^a=e^a_\sigma V^\sigma$ on the RHS
\begin{eqnarray}
    \nabla_\mu V^\nu
        &=&e_a^\nu \partial_\mu (e_\sigma^a V^\sigma)+e_a^\nu e^b_\sigma \omega_{\mu\phantom a b}^{\phantom \mu a} V^\sigma\\ 
        &=& (e_a^\nu \partial_\mu e_\sigma^a) V^\sigma+e_a^\nu e_\sigma^a \partial_\mu V^\sigma+ e_a^\nu e_\sigma^b \omega_{\mu\phantom a b}^{\phantom \mu a}V ^\sigma
\end{eqnarray}
and recall that $e^\nu_a e^a_\sigma = \delta^\nu_\sigma$ and $e_a^\nu e_\sigma^b \omega_{\mu\phantom a b}^{\phantom \mu a}=\omega_{\mu\phantom\nu\sigma}^{\phantom \mu\nu}$, where $\omega_{\mu\phantom\nu\sigma}^{\phantom \mu\nu}$ is given by \eqref{spin-connection}. In this case
\begin{eqnarray}
    \nabla_\mu V^\nu &=& (e_a^\nu \partial_\mu e_\sigma^a)V^\sigma+\partial_\mu V^\nu+(\Gamma_{\mu\sigma}^{\nu}-e_a^\nu \partial_\mu e_\sigma^a)V^\sigma.
\end{eqnarray}
The terms with $e_a^\nu \partial_\mu e_\sigma^a$ cancel and we are left with
\begin{eqnarray}\nabla_\mu V^\nu =\partial_\mu V^\nu+\Gamma^\nu_{\mu\sigma}V^\sigma,
\end{eqnarray} which agrees with the Levi-Civita covariant derivative of the vector field with respect to the coordinate components. 

\section{Conformal Killing vector field action in the strip}
\label{app:ckv-action}

The components of the rotation generators with respect to the vielbein
\begin{eqnarray}
    e^1 = \sqrt{\dfrac{\gamma_{z\bar z}}{2}}(dz+d\bar z),\quad e^2 =-i\sqrt{\dfrac{\gamma_{z\bar z}}{2}}(dz-d\bar z),\quad e^3 = \dfrac{du}{R}
\end{eqnarray}
 are $S_{\mu\nu} = e^a_{\mu} e^b_{\nu} S_{ab}$. Explicitly, we find
\begin{eqnarray}
    S_{uz} &=&\dfrac{i}{R}\sqrt{\frac{\gamma_{z\bar z}}{2}}J_-\label{eq:cyl-spin-gen-1},\\
    S_{u\bar z} &=&-\dfrac{i}{R}\sqrt{\frac{\gamma_{z\bar z}}{2}}J_+\label{eq:cyl-spin-gen-2},\\
    S_{z\bar z} &=&i\gamma_{z\bar z}J_3\label{eq:cyl-spin-gen-3},
\end{eqnarray}
where 
\be 
J_- = S_{23} - i S_{31}, \quad J_+ = S_{23} + i S_{31}, \quad J_3 = S_{12}.
\ee

The coordinate components of the torsion-free spin connection $\omega_{\mu\phantom\sigma\nu}^{\phantom \mu\sigma}$ are given by
\begin{eqnarray}
\label{spin-connection}
    \omega_{\mu\phantom\sigma\nu}^{\phantom\mu\sigma} = \Gamma^\sigma_{\mu\nu}-e^\sigma_a \partial_\mu e^a_\nu
\end{eqnarray}
and therefore, we see that its only non-vanishing components are
\begin{eqnarray}
    \omega_{z\phantom z z}^{\phantom zz} = -\omega_{z\phantom {\bar z}\bar z}^{\phantom z \bar z}= \frac{1}{2}\Gamma^z_{zz},\\
    \omega_{\bar z\phantom {\bar z} \bar z}^{\phantom {\bar z}\bar z} = - \omega_{\bar z\phantom {z} z}^{\phantom {\bar z} z} = \frac{1}{2}\Gamma^{\bar z}_{\bar z\bar z},
\end{eqnarray}
 where
\be 
\Gamma^{z}_{~zz} = -\frac{2\bz}{1 + z\bz}, \quad \Gamma^{\bz}_{~\bz\bz} = -\frac{2z}{1 + z\bz}.
\ee 
As a result, defining
\begin{eqnarray}
    \Omega_z &\equiv& \dfrac{1}{2}\Gamma^z_{zz},\quad \Omega_{\bar z} \equiv -\dfrac{1}{2}\Gamma^{\bar z}_{\bar z \bar z}
\end{eqnarray}
we find that the spin covariant derivative of ${\cal O}_\Delta$ is given by
\begin{eqnarray}
    \nabla_u{\cal O}_\Delta &=& \partial_u{\cal O}_\Delta,\label{eq:cyl-cov-der-1}\\
    \nabla_z{\cal O}_\Delta &=& \partial_z{\cal O}_\Delta - \Omega_z J_3{\cal O}_\Delta,\label{eq:cyl-cov-der-2}\\
    \nabla_{\bar z}{\cal O}_\Delta &=& \partial_{\bar z}{\cal O}_{\Delta} -\Omega_{\bar z}J_3{\cal O}_\Delta.\label{eq:cyl-cov-der-3}
\end{eqnarray}

Now fix $\tau_0 = \frac{\pi}{2}$ and take $\epsilon = L_Y$ given by
\begin{eqnarray}
    L_Y &\equiv& \dfrac{\epsilon_Y^+-\epsilon_Y^-}{2}\label{eq:defn-epsilon_Y}\\
    &=&\frac{i}{2}(D\cdot Y) u\partial_u + i Y^A\partial_A+O(R^{-1}).
\end{eqnarray}
We will show that ${\delta}_{L_Y}{\cal O}_{\Delta}$ reproduces the action of $Y$ on a 2D primary operator in the large $R$ limit. To this end observe from \eqref{eq:cyl-cov-der-1}-\eqref{eq:cyl-cov-der-3} and \eqref{eq:cyl-spin-gen-1}-\eqref{eq:cyl-spin-gen-3} that for this vector field we have 
\begin{eqnarray}
    \nabla\cdot {L}_Y &=& i\frac{3}{2}D\cdot Y + \mathcal{O}(R^{-1}),\\
    {L}_Y^\mu \nabla_\mu \mathcal{O}_{\Delta} &=& i\left[\frac{1}{2}D\cdot Y u\partial_u+Y^z(\partial_z -\Omega_z J_3)+Y^{\bar z}(\partial_{\bar z} - \Omega_{\bar z}J_3)+ O(R^{-1}) \right] \mathcal{O}_{\Delta},\\
    \frac{i}{2}\nabla_{\mu}({L}_{Y})_\nu S^{\mu\nu} &=& \dfrac{i}{2}(D_z Y^z- D_{\bz}Y^{\bz})J_3+ O(R^{-1}).
\end{eqnarray}
From this we immediately see that the expansion of $\delta_{{L}_Y}{\mathcal{O}_{\Delta}}(x)$ is
\begin{equation}
    \delta_{{L}_Y}{\cal O}_\Delta(x) = - i \left[D_zY^z \mathfrak{h} +D_{\bar z}Y^{\bar z}\bar{\mathfrak{ h}}+Y^z(\partial_z -\Omega_z J_3)+Y^{\bar z}(\partial_{\bar z} -\Omega_{\bar z}J_3)+O(R^{-1})\right]{\cal O}_{\Delta}(x).  
\end{equation}
Here we have defined the operator-valued weights
\begin{eqnarray}
    \mathfrak{h} \equiv \dfrac{\hat{\Delta}+J_3}{2},\quad \bar{ \mathfrak{h}} \equiv\dfrac{\hat{\Delta}-J_3}{2},\quad \hat{\Delta} \equiv  \Delta+u\partial_u.
\end{eqnarray}
This agrees precisely with the transformation of a 2D primary operator, as given for example in \cite{Kapec:2016jld}.

\section{Shadows and dimensional reduction}
\label{app:shadow-dim-red}

In this appendix we discuss the connection between the $d$-dimensional shadow transform on the cylinder and the Mellin-like transform on an infinitesimal time strip that implements the dimensional reduction to $S^{d-1}$. All embedding space fields are assumed to obey the properties described in appendix \ref{sec:embedding-space}. We begin by projecting the embedding space formula for the shadow transform to a particular section. Starting from \eqref{eq:embedding-shadow}, we find
\begin{equation}
\begin{split}
\widetilde{\Phi}_{\mu_1\cdots \mu_J}(x)&= \prod_{i}\dfrac{\partial P^{A_i}}{\partial x^{\mu_i}}\widetilde{\Phi}_{A_1\cdots A_J}(P(x)) \\
&= \prod_{i}\dfrac{\partial P^{A_i}}{\partial x^{\mu_i}}\int D^d P(y)\dfrac{\prod_{i}(\eta_{A_iB_i} P(x)\cdot P(y) -P_{A_i}(y)P_{B_i}(x))}{(-2P(x)\cdot P(y))^{d-\Delta+J}}\prod_{i}\eta^{B_iC_i}{\Phi}_{C_1\cdots C_J}(P(y)),\\
\end{split}
\end{equation}
where the conformal integral is gauge-fixed to a particular section $Y = P(y)$. We now use \eqref{eq:completeness-relation} to eliminate $\eta^{B_iC_i}$, noting that the  $q^{(B_i}\bar q^{C_i)}$ contributions contract to zero, namely
\begin{eqnarray}
\widetilde{\Phi}_{\mu_1\cdots \mu_J}(x)
&=&\prod_{i}\dfrac{\partial P^{A_i}}{\partial x^{\mu_i}}\int D^d P(y)\dfrac{\prod_{i}(\eta_{A_iB_i} P(x)\cdot P(y) -P_{A_i}(y)P_{B_i}(x))}{(-2P(x)\cdot P(y))^{d-\Delta+J}}\nonumber\\&&\times\prod_{i}g^{\sigma_i\rho_i}(y)\dfrac{\partial P^{B_i}}{\partial y^{\sigma_i}}\dfrac{\partial P^{C_i}}{\partial y^{\rho_i}}{\Phi}_{C_1\cdots C_J}(P(y))\\
&=&\int D^d P(y)\dfrac{\prod_{i}\frac{\partial P^{A_i}}{\partial x^{\mu_i}}\frac{\partial P^{B_i}}{\partial y^{\nu_i}}(\eta_{A_iB_i} P(x)\cdot P(y) -P_{A_i}(y)P_{B_i}(x))}{(-2P(x)\cdot P(y))^{d-\Delta+J}}{\Phi}^{\nu_1\cdots \nu_J}(y).
\end{eqnarray}
We finally observe that owing to \eqref{eq:inversion-tensor-embedding} we can write 
\begin{eqnarray}
\widetilde{\Phi}_{\mu_1\cdots \mu_J}(x)
&=&\int d^dy \sqrt{g(y)}\dfrac{\prod_{i}\partial_{x^{\mu_i}}\partial_{y^{\nu_i}}\log(-2P(x)\cdot P(y))}{(-2P(x)\cdot P(y))^{d-\Delta}}{\Phi}^{\nu_1\cdots \nu_J}(y),
\end{eqnarray}
which is the shadow transform restricted to a section of lightcone \cite{Simmons-Duffin:2012juh}. 

Now we consider the particular case of the cylinder section parameterized by \eqref{cylinder-embedding} and expand at large $R$. In this case taking $x = (\tau,\Omega)$ and $y = (\tau',\Omega')$ we have
\begin{eqnarray}
	P(x) \cdot P(y) 
	&=& - \cos(\tau-\tau') + \Omega\cdot \Omega'.  
\end{eqnarray}
Setting $\tau = \pm \frac{\pi}{2}+ \frac{u}{R}$, expanding at large $R$ and taking the time Mellin-like transform \eqref{eq:time-Mellin} we find
\be 
\label{shadow-Mellin0}
\begin{split}
&\Gamma(\Delta_0)\int_{-\infty}^{\infty} du u^{-\Delta_0}\widetilde{\Phi}^{\pm}_{\mu_1 \cdots \mu_J}(u, \Omega) = \Gamma(\Delta_0)\int_{-\infty}^{\infty} du u^{-\Delta_0}\\
&\times \int d\tau' d^{d-1}\vec{z}^{~'} \dfrac{\prod_{i}\partial_{x^{\mu_i}}\partial_{y^{\nu_i}}\log(\pm 2 \sin \tau' \mp 2\frac{u}{R}\cos \tau' - 2 \Omega \cdot \Omega')}{(\pm 2 \sin \tau' \mp 2\frac{u}{R}\cos \tau' - 2 \Omega \cdot \Omega')^{d-\Delta}}{\Phi}^{\nu_1\cdots \nu_J}(y)\\
&= -i \frac{\Gamma(\Delta_0)}{\Gamma(d - \Delta)}\int_{-\infty}^{\infty} du u^{-\Delta_0} \int d\tau' d^{d-1}\vec{z}^{~'} \int_0^{\infty} d\omega (-i\omega)^{d - \Delta- 1} e^{i \omega (\pm 2 \sin \tau' \mp 2\frac{u}{R}\cos \tau' - 2 \Omega \cdot \Omega')}\\
&\times F_{\mu_1 \cdots \mu_i}(x, y),
\end{split}
\ee
where 
\be 
F_{\mu_1 \cdots \mu_i}(x, y) = \prod_{i}\partial_{x^{\mu_i}}\partial_{y^{\nu_i}}\log(\pm 2 \sin \tau' -2 \Omega \cdot \Omega') {\Phi}^{\nu_1\cdots \nu_J}(y) + \mathcal{O}(R^{-1})
\ee
and $\mu_i, \nu_i$ are restricted to $\Omega, \Omega'$. We also defined 
\be 
\Phi^{\pm}(u, \Omega) \equiv \Phi(\pm \frac{\pi}{2} + \frac{u}{R}, \Omega).
\ee
In general, $\int du u^{-\Delta_0}\widetilde{\Phi}$ is an operator in CFT$_d$ with dimension $d - \Delta + \Delta_0 - 1$ (see section \ref{sec:infinite-dimensional}). Setting $\Delta_0 = 0$ should then yield an operator of dimension $d - \Delta - 1$ in CCFT$_{d-1}$. Note that for $\Delta_0 = 0$, \eqref{shadow-Mellin0} is singular which suggests one should take a residue \cite{Pate:2019mfs}. Indeed, the residue of \eqref{shadow-Mellin0} at $\Delta_0 = 0$ reduces to
\be 
\begin{split}
\int_{-\infty}^{\infty} du \widetilde{\Phi}^{\pm}_{\mu_1 \cdots \mu_J}(u, \Omega) &= - \frac{1}{\Gamma(d - \Delta)} \int d\tau' d^{d-1}\vec{z}^{~'} \int_0^{\infty} d\omega (-i\omega)^{d -1 - \Delta- 1} \frac{R}{2}\sum_{\tau_0 = \pm \frac{\pi}{2}}\delta(\tau'  - \tau_0) \\
&\times  e^{i \omega (\pm 2 \sin \tau' - 2\Omega \cdot \Omega')} F_{\mu_1 \cdots \mu_i}(x, y) \\
&= -\frac{i}{2} \frac{R}{d - 1 - \Delta}\int d^{d-1}\vec{z}^{~'} \sum_{\alpha \in \{0,1 \}}\frac{\prod_{i}\partial_{x^{\mu_i}}\partial_{y^{\nu_i}}\log(\pm e^{i\pi \alpha} 2 - 2 \Omega \cdot \Omega') }{(\pm e^{i\pi \alpha} 2 -2 \Omega \cdot \Omega')^{d - 1 - \Delta}} \\
&\times \Phi^{\nu_1 \cdots \nu_J}(e^{i\pi \alpha} \frac{\pi}{2}, \Omega') + \mathcal{O}(R^{0}),
\end{split}
\ee
which we recognize as proportional to a linear combination of $(d-1)$-dimensional shadow transforms in the strips around $\pm \frac{\pi}{2}$. Note the appearance of a  linear combination of incoming and outgoing insertions. It may be interesting to understand this better, perhaps in relation to the proposal of \cite{Crawley:2021ivb}.

On the other hand, taking the residue at $\Delta_0 = 1$ of  \eqref{shadow-Mellin0} and using the identity \cite{Pate:2019mfs}
\be 
\lim_{\epsilon \rightarrow 0} \epsilon x^{\epsilon - 1} = 2\delta(x),
\ee 
we find 
\be 
\underset{\Delta_0 = 1}{\rm Res} {\Gamma(\Delta_0)}\int_{-\infty}^{\infty} du u^{-\Delta_0}\widetilde{\Phi}^{\pm}_{\mu_1 \cdots \mu_J}(u, \Omega) = 2 \widetilde{\Phi}^{\pm}_{\mu_1 \cdots \mu_J}(0, \Omega).
\ee
This operator is a primary of dimension $d - \Delta$ in the CFT$_d$ as well as in the CCFT$_{d-1}$. For $d = 3$, taking a 2D shadow then yields an operator of dimension $\Delta - 1$, which in the special case of the CFT$_3$ stress tensor reduces to the stress tensor in the CCFT$_2$. 

 More generally, given operators ${\cal O}^{\pm}_\Delta(u,\Omega)$  in strips around $\pm \frac{\pi}{2}$,
\be 
\underset{\Delta_0 = 1}{\rm Res}\int_{-\infty}^{\infty} du u^{-\Delta_0}{\cal O}^{\pm}_\Delta(u, \Omega)=2{\cal O}^{\pm}_\Delta(0,\Omega). 
\ee 
Since ${\Delta}_{\rm CCFT} = \Delta+\Delta_0-1$ we get an operator of $\Delta_{\rm CCFT} =\Delta$. We conclude that placing an operator at $u = 0$ inside a small time interval corresponds in CCFT to an operator that inherits the  dimension $\Delta$ of the operator in the parent CFT, as found in \cite{PipolodeGioia:2022exe}.

\section{Derivation of CCFT$_{d-1}$ conformally soft theorems from CFT$_d$}\label{app:soft-graviton-theorems}

In this appendix, we give the derivation of the leading and subleading conformally soft graviton theorems from the higher dimensional shadow stress tensor correlator. We start by defining
\begin{eqnarray}
    S_{ab}^{(d)} &=&\sum_{i=1}^n \dfrac{\varepsilon_{a}^{A}\varepsilon_b^B(x) P_A(x_i) P^C(x)}{P(x)\cdot P(x_i)}({\cal J}_i)_{CB},
\end{eqnarray}
so that the shadow stress tensor correlator in the CFT$_{d}$ becomes
\begin{eqnarray}
    \langle {\mathscr G}_{ab}{\cal O}_1\cdots{\cal O}_n\rangle &=& -i S_{\{ab\}}^{(d)}\langle{\cal O}_1\cdots{\cal O}_n\rangle.
\end{eqnarray}
To compute the flat space limit of $S_{ab}^{(d)}$ we expand at large $R$ keeping the first subleading contributions. To keep track of them we introduce the following notation:
\begin{eqnarray}
    P &=& q + \delta q,\quad \varepsilon_a = \epsilon_a + \delta \epsilon_a,\quad a\in \{1,\dots, d-1\},
\end{eqnarray}
where $q = (q^0,q^i,0)$ denotes the leading term in $P$ and $\epsilon_a = (\epsilon_a^0,\epsilon_a^i,0)$ the leading term in $\varepsilon_a$. These correspond to the flat space counterparts of $P$ and $\varepsilon_a$. $\delta q$ and $\delta \epsilon_a$ are the deviations from the flat space limit and take the form
\begin{eqnarray}
    \delta q &=& (\sin\tau-1,\vec{0},\cos\tau),\quad \delta \epsilon_a = z_a \delta q, \quad a \in \{1, \cdots, d - 1 \}.
\end{eqnarray}
We restrict our attention to the components of the shadow stress tensor tangent to the $S^{d-1}$ on which the CCFT is defined, namely with $a \in \{1, \cdots, d - 1 \}$. 

We need to evaluate
\begin{eqnarray}
    \dfrac{\varepsilon_a(x)\cdot P(x_i)}{P(x)\cdot P(x_i)},\quad P^A(x)\varepsilon_b^B(x)({\cal J}_i)_{AB}.
\end{eqnarray}
The first quantity is immediate to expand and yields
\begin{eqnarray}\label{eq:soft-factor-1}
    \dfrac{\varepsilon_a(x)\cdot P(x_i)}{P(x)\cdot P(x_i)} 
    &=&\dfrac{\epsilon_a(x)\cdot q(x_i)}{q(x)\cdot q(x_i)}+O(R^{-1}).
\end{eqnarray}
For the second one we have
\begin{eqnarray}
    P^A(x)\varepsilon_b^B(x)({\cal J}_i)_{AB} 
    &=&q^A(x)\epsilon_b^B(x)({\cal J}_i)_{AB}+z_bq^A(x)\delta q^B(x)({\cal J}_i)_{AB}+\delta q^A(x)\epsilon_b^B(x)({\cal J}_i)_{AB}.\nonumber\\
\end{eqnarray}
We now study the second and third terms observing that for and $\tau = \frac{\pi}{2} + \frac{u}{R}$ and large $R$, $({\cal J}_i)_{A,d+1} = i Rq_A(x_i)\partial_{u_i} + O(1)$ 
\begin{eqnarray}
    q^A(x)\delta q^B(x)({\cal J}_i)_{AB} 
    &=&-(\sin\tau-1)q^j(x)({\cal J}_i)_{0j} + \cos\tau q(x)\cdot q(x_i) \left(iR\partial_{u_i} + O(R^0)\right),\\
    \delta q^A(x)\epsilon_b^B(x)({\cal J}_i)_{AB} 
    &=&(\sin\tau-1)\epsilon_b^j(x)({\cal J}_i)_{0j}-\cos\tau \epsilon_b(x)\cdot q(x_i) \left(iR\partial_{u_i} + O(R^0)\right).
\end{eqnarray}
As a result, we have
\begin{eqnarray}
    P^A(x)\varepsilon_b^B(x)({\cal J}_i)_{AB} &=& q^A(x)\epsilon_b^B(x)({\cal J}_i)_{AB}- z_b (\sin\tau-1)q^j(x)({\cal J}_i)_{0j} \nonumber\\
    &+&z_b \cos\tau q(x)\cdot q(x_i) \left(i R\partial_{u_i} + O(R^0)\right)
    +(\sin\tau-1)\epsilon_b^j(x)({\cal J}_i)_{0j}\nonumber\\
    &-& \cos\tau \epsilon_b(x)\cdot q(x_i) \left(i R\partial_{u_i} + O(R^0)\right).
\end{eqnarray}
At this point, we can further expand at large $R$. In particular, we notice that the first term is $O(1)$ because $A,B<d+1$. For the others we write $\tau = \frac{\pi}{2}+\frac{u}{R}$ and expand at large $R$ to find
\begin{eqnarray}
    P^A(x)\varepsilon_b^B(x)({\cal J}_i)_{AB} &=& q^A(x)\epsilon_b^B(x)({\cal J}_i)_{AB}- i uz_b q(x)\cdot q(x_i)\partial_{u_i} + i u\epsilon_b(x)\cdot q(x_i)\partial_{u_i}\nonumber\\
    &&+O(R^{-1}).
\end{eqnarray}
Combining with \eqref{eq:soft-factor-1} we find
\begin{equation}
\label{eq:final-expansion}
\begin{split}
     S_{ab}^{(d)}
     &=\sum_{i=1}^n \dfrac{\varepsilon_a(x)\cdot P(x_i)}{P(x)\cdot P(x_i)}P^A(x)\varepsilon_b^B(x)({\cal J}_i)_{AB}\\
    &=\sum_{i=1}^n \bigg[ \dfrac{\epsilon_a(x)\cdot q(x_i)}{q(x)\cdot q(x_i)} \bigg(q^A(x)\epsilon_b^B(x)({\cal J}_i)_{AB}- i uz_b q(x)\cdot q(x_i)\partial_{u_i} + i u\epsilon_b(x)\cdot q(x_i)\partial_{u_i}\bigg)+O(R^{-1})\bigg].\\
    \end{split}
\end{equation}

Taking one derivative in $u$ we get
\begin{eqnarray}
     \partial_u S_{ab}^{(d)}&=& i\sum_{i=1}^n \bigg[\dfrac{\epsilon_a(x)\cdot q(x_i)}{q(x)\cdot q(x_i)}\bigg(-z_b q(x)\cdot q(x_i)\partial_{u_i} + \epsilon_b(x)\cdot q(x_i)\partial_{u_i}\bigg)+O(R^{-1})\bigg]\nonumber\\
     &=& i\sum_{i=1}^n \bigg[\bigg(-z_b\epsilon_a(x)\cdot q(x_i)\partial_{u_i} + \dfrac{\epsilon_a(x)\cdot q(x_i)\epsilon_b(x)\cdot q(x_i)}{q(x)\cdot q(x_i)}\partial_{u_i}\bigg)+O(R^{-1})\bigg].
\end{eqnarray}
Now observe that the first term is proportional to the operator $\sum_{i} q^A(x_i)\partial_{u_i}$ which annihilates conformal correlators by the global conformal symmetry of the CFT$_d$ to leading order at large $R$ (or equivalently by momentum conservation in the flat limit). Specifically 
\be 
\sum_{i = 1}^n \mathcal{J}_{j, d+1}(x_i)\langle \mathbb{X}\rangle = \sum_{i = 1}^n \left(-i P_j(x_i) \p_{P^{d+1}(x_i)} +i P_{d+1}(x_i) \p_{P^{j}(x_i)} + \mathcal{S}_{j,d+1} \right) \langle \mathbb{X}\rangle = 0, \quad j = 0, \cdots d,
\ee
and therefore
\be 
\sum_{i=1}^n i q_j(x_i) \p_{u_i} \langle \mathbb{X} \rangle = \frac{1}{R} \sum_{i=1}^n \left(-i P_{d+1}(x_i) \p_{P^j (x_i)} - \mathcal{S}_{j,d+1} \right)\langle \mathbb{X} \rangle = \mathcal{O}(R^{-1}), \quad j = 0, \cdots d.
\ee
As such, only the second term remains
\begin{eqnarray}
     \partial_u S_{ab}^{(d)}
     &=& i\sum_{i=1}^n \bigg[\dfrac{\epsilon_a(x)\cdot q(x_i)\epsilon_b(x)\cdot q(x_i)}{q(x)\cdot q(x_i)}\partial_{u_i}+O(R^{-1})\bigg],
\end{eqnarray}
which coincides with the leading soft factor. Moreover, it is also clear that
\begin{eqnarray}
    (1-u\partial_u)S_{ab}^{(d)} &=&\sum_{i=1}^n \dfrac{\epsilon_a(x)\cdot q(x_i)}{q(x)\cdot q(x_i)}q^A(x)\epsilon_b^B(x)({\cal J}_i)_{AB}+O(R^{-1}),
\end{eqnarray}
where since $a,b\in \{1,\dots,d-1\}$ it follows that $A,B \in \{0,\dots,d\}$ and in this range $({\cal J}_i)_{AB}$ act as the $\mathbb{R}^{1,d}$ Lorentz generators in the flat space limit. Finally we take the $(d-1)$-dimensional symmetric traceless component of $S_{ab}^{(d)}$ with  $a,b\in \{1,\dots,d-1\}$ by applying the projector \eqref{eq:projector}. Then
\begin{eqnarray}
   \epsilon_{ab} \equiv \epsilon_{\{a}^A\epsilon^B_{b\}} &=& \dfrac{1}{2}[\epsilon_a^A\epsilon_b^B+\epsilon_a^B\epsilon_b^A]-\dfrac{\eta_{ab}}{d -1}[\eta^{cd}\epsilon_c^A\epsilon_d^B].
\end{eqnarray}
However, since $\epsilon^{d+1}_a =0$ it follows that $\eta^{cd}\epsilon_c^A\epsilon_d^B = \delta^{cd}\epsilon_c^A\epsilon_d^B$ and that $\epsilon_{\{a}^A\epsilon^B_{b\}}=0$ when either $A$ or $B$ are $d+1$. As a result, for $a,b\in \{1,\dots,d-1\}$,
\begin{eqnarray}
    \epsilon_{\{a}^A\epsilon^B_{b\}} &=& \dfrac{1}{2}[\epsilon_a^A\epsilon_b^B+\epsilon_a^B\epsilon_b^A]-\dfrac{\delta_{ab}}{d -1}[\delta^{cd}\epsilon_c^A\epsilon_d^B],\quad A,B<d+1,
\end{eqnarray}
which coincide with the symmetric traceless polarizations in $\mathbb{R}^{1,d}$. As a result, the operators $\mathscr{N}_{ab}^{(0)} =  \underset{u\rightarrow 0}{\rm lim} \partial_u \mathscr{G}_{{\{ab\}}}$ and $\mathscr{N}_{ab}^{(1)}= \underset{u\rightarrow 0}{\rm lim}(1-u\p_u)\mathscr{G}_{{\{ab\}}}$ play the role of leading and subleading conformally soft gravitons in $\mathbb{R}^{1, d}$. It is immediate to see that they have the expected dimensions $\Delta=1$ and $\Delta=0$ respectively.

 We conclude this appendix with a comment on the timelike components of the shadow stress tensor. For $d = 3$ one can construct from the $u, A$ components of the shadow stress tensor operators which coincide with the supertranslation currents in the dimensionally reduced theory. This is perhaps to be expected, as conservation of the CFT$_3$ stress tensor leads to relations among its transverse and time components. It may be interesting to further explore these constraints in relation to the asymptotic Einstein equations in $4$D AFS.

\section{Subleading soft factor in CCFT$_2$}
\label{app:subleading-soft-factor-ccft2}

In this appendix we calculate the subleading soft factor
\begin{eqnarray}
    (1-u\partial_u)S_{ab}^{(d)} &=&\sum_{i=1}^n \dfrac{\epsilon_a(x)\cdot q(x_i)}{q(x)\cdot q(x_i)}q^A(x)\epsilon_b^B(x)({\cal J}_i)_{AB}+O(R^{-1}),
\end{eqnarray}
in the specific case of reduction from CFT$_3$ to CCFT$_2$. We need to evaluate $q^A(x)\epsilon_b^B(x)({\cal J}_i)_{AB}$ using the complex polarization vectors $\{\epsilon_z,\epsilon_{\bz}\}$. We recall that $({\cal J}_i)_{AB}$ are the $\mathfrak{so}(3,2)$ generators acting on the $i$-th primary operator. The actions of such conformal Killing vectors and their large $R$ expansion have been studied in section \ref{sec:ckv-action}. In particular, we note that since $q^{4}=\epsilon_b^{4}=0$, only $({\cal J}_i)_{AB}$ with $A,B< 4$ appear. For this range of indices, we have\footnote{It is possible to check by explicit computation that ${\cal J}_{AB}$ reproduces the conformal Killing vector action by studying its action on lightcone fields in coordinates adapted to the cylinder section. Indeed, parameterizing the lightcone as $X = (r\sin\tau,r\Omega,r\cos\tau)$, so that the cylinder section is obtained by gauge-fixing $r=1$, and evaluating ${\cal J}_{AB}{\cal O}_\Delta(X)$, we find due to the homogeneity of ${\cal O}_\Delta(X)$ under rescalings that $-r\p_r{\cal O}_\Delta = \Delta{\cal O}_\Delta$. Then \eqref{eq:embedding-generators-ckv} follows by straightforward computation.} 
\begin{eqnarray}\label{eq:embedding-generators-ckv}
    {\cal J}_{AB}{\mathcal{O}_i} &=& -\delta_{L_{Y_{AB}}}{\mathcal{O}_i}, \quad A, B = 0, \cdots 3,
\end{eqnarray}
where $L_{Y}$ has been defined in \eqref{eq:lorentz-generators} and $Y_{AB}$ are the $S^2$ conformal Killing vectors \eqref{eq:rot-sph} and \eqref{eq:boosts}. We have computed the large $R$ expansion of $\delta_{L_{Y_{AB}}}{\cal O}_i$ in \eqref{eq:ckv-action-flat-limit-1}, which yields 
\begin{equation}
    ({\cal J}_i)_{AB}{\mathcal{O}_i} = i \left( D_{z_i}Y^{z_i}_{AB} \mathfrak{h}_{i} +D_{\bz_i}Y^{\bz_i}_{AB}\bar{\mathfrak{ h}}_i+Y^{z_i}_{AB}(\partial_{z_i} - \Omega_{z_i} J_3)+Y^{\bz_i}_{AB}(\partial_{\bz_i} - \Omega_{\bz_i}J_3)+O(R^{-1}) \right)\mathcal{O}_i.
\end{equation}
Now using the explicit parametrization of $q$ and $\{\epsilon_z,\epsilon_{\bz}\}$ it is straightforward to compute the following contractions 
\begin{eqnarray}
    q^A(x)\epsilon^B_{\bz}(x)Y_{AB}(z_i,\bz _i) &=& -{\color{red}}\dfrac{(z-z_i)^2}{1+z\bz}\partial_{z_i},\\
    q^A(x)\epsilon^B_{z}(x)Y_{AB}(z_i,\bz _i)&=&-{\color{red}}\dfrac{(\bz -\bz_i)^2}{1+z\bz}\partial_{\bz_i},
\end{eqnarray}
from which we immediately obtain 
\begin{equation}
\begin{split}
  - i q^A(x)\epsilon_{\bz}^B(x)({\cal J}_i)_{AB}\mathcal{O}_i &= \left[ \dfrac{(z-z_i)(1+z \bar z_i)}{(1+z\bar z)(1+z_i\bar z_i)}2{\mathfrak{h}}_i-\dfrac{(z- z_i)^2}{1+z\bar z}(\partial_{z_i}{-}\Omega_{z_i}J_3)+O(R^{-1}) \right]\mathcal{O}_i,\\
   -  i q^A(x)\epsilon_z^B(x)({\cal J}_i)_{AB}\mathcal{O}_i &= \left[\dfrac{(\bar z-\bar z_i)(1+\bar z z_i)}{(1+z\bar z)(1+z_i\bar z_i)}2\bar{\mathfrak{h}}_i-\dfrac{(\bar z-\bar z_i)^2}{1+z\bar z}(\partial_{\bz_i}{-}\Omega_{\bz_i}J_3)+O(R^{-1})\right]\mathcal{O}_i.
    \end{split}
\end{equation}
In turn, this means that we have 
\begin{eqnarray}
    (1-u\partial_u)S_{\bar z \bar z}^{(3)} &=& i\sum_{i=1}^n \left[\dfrac{(z-z_i)(1+z \bar z_i)}{(\bar z-\bar z_i)(1+z_i\bar z_i)}2{\mathfrak{h}}_i-\dfrac{(z- z_i)^2}{\bar z-\bar z_i}(\partial_{z_i}{-}\Omega_{z_i}J_3)\right]+O(R^{-1}),\\
    (1-u\partial_u)S_{zz}^{(3)} &=& i \sum_{i=1}^n \left[\dfrac{(\bz-\bz_i)(1+\bz z_i)}{( z- z_i)(1+z_i\bar z_i)}2\bar{\mathfrak{h}}_i-\dfrac{(\bz- \bz_i)^2}{ z- z_i}(\partial_{\bz_i}{-}\Omega_{\bz_i}J_3)\right]+O(R^{-1}),
\end{eqnarray}
which take the form of the standard CCFT$_2$ soft factors \cite{Kapec:2014opa, Kapec:2016jld} with the operator-valued weights $(\mathfrak{h},\bar{\mathfrak{h}})$ in place of the standard weights.

\bibliographystyle{utphys}
\bibliography{references}

\end{document}